\documentclass[prd,twocolumn,showpacs,floatfix,amsmath,amssymb,floatfix]{revtex4}
\usepackage{graphicx,color,dcolumn,booktabs,bm}
\usepackage{longtable,lscape}
\usepackage{txfonts}
\usepackage{overpic}
\usepackage{amssymb}
\usepackage{indentfirst}
\usepackage{feynmf}   %{feynmp}
\usepackage{slashed}  %for Feynman symbols
\usepackage{cases}
\usepackage{color}
\usepackage{multirow}
\usepackage{graphicx,color,dcolumn,booktabs,bm}
\usepackage[dvipdfm,  %pdftex,pdflatex
            colorlinks, % (\colorlinks\pdfborder)
            pdfborder=001,   %
            citecolor=blue
            ]{hyperref}

\graphicspath{{Figures/}} %

\begin{document}
%\begin{CJK}{GBK}{}

\title{Higher bottom and bottom-strange mesons}
\author{Yuan Sun$^{1,2}$}
\author{Qin-Tao Song$^{1,3,6}$}
\author{Dian-Yong Chen$^{1,3}$}
\author{Xiang Liu$^{1,2}$\footnote{Corresponding author}}\email{xiangliu@lzu.edu.cn}
\author{Shi-Lin Zhu$^{4,5}$\footnote{Corresponding author}}
\email{zhusl@pku.edu.cn} \affiliation{$^1$Research Center for Hadron
and CSR Physics, Lanzhou University $\&$ Institute of Modern Physics
of CAS,
Lanzhou 730000, China\\
$^2$School of Physical Science and Technology, Lanzhou University,
Lanzhou 730000, China\\
$^3$Nuclear Theory Group, Institute of Modern Physics of CAS,
Lanzhou 730000, China\\
$^4$Department of Physics and State Key Laboratory of Nuclear
Physics and Technology, Peking University, Beijing 100871, China\\
$^5$Collaborative Innovation Center of Quantum Matter, Beijing
100871, China\\
$^6$University of Chinese Academy of Sciences, Beijing 100049, China}

\begin{abstract}

Motivated by the recent observation of the orbital excitation
$B(5970)$ by the CDF Collaboration, we have performed a systematic
study of the mass spectrum and strong decay patterns of the higher
$B$ and $B_s$ mesons. Hopefully the present investigation may
provide valuable clues to further experimental exploration of these
intriguing excited heavy mesons.

\end{abstract}
 \maketitle

\section{introduction}\label{sec1}

The past decade has witnessed the discovery of many charmed and
charmed-strange states such as $D_{sJ}(2317)$
\cite{Aubert:2003fg,Besson:2003cp,Krokovny:2003zq}, $D_s(2460)$
\cite{Besson:2003cp,Krokovny:2003zq}, $D_{sJ}(2632)$
\cite{Evdokimov:2004iy}, $D_{sJ}(2860)$ \cite{Aubert:2006mh},
$D_{sJ}(2715)$ \cite{Abe:2006xm}, $D_{sJ}(3040)$
\cite{Aubert:2009ah}, $D(2550)$
\cite{delAmoSanchez:2010vq}/$D_J(2580)$ \cite{Aaij:2013sza},
$D^*(2600)$ \cite{delAmoSanchez:2010vq}/$D_J^*(2650)$
\cite{Aaij:2013sza}, $D(2750)$
\cite{delAmoSanchez:2010vq}/$D_J(2740)$ \cite{Aaij:2013sza},
$D^*(2760)$ \cite{delAmoSanchez:2010vq}/$D_J^*(2760)$
\cite{Aaij:2013sza}, and $D_J^{(*)}(3000)$ \cite{Aaij:2013sza}, which
have not only enriched the family of the charmed and charmed-strange
states, but also stimulated extensive discussions of their
properties (see Ref. \cite{Liu:2010zb} for a mini review of the
research status of these newly observed charmed and charmed-strange
states).

The present situation of the experimental exploration of bottom and
bottom-strange states is strikingly similar to that of charmed and
charmed-strange states in 2003; i.e., some candidates for the
P-wave bottom and bottom-strange meson were announced
\cite{Abreu:1994hj,Akers:1994fz,Buskulic:1995mt,Barate:1998cq,Abazov:2007vq,Aaltonen:2008aa,Aaltonen:2007ah,Abazov:2007af,Aaij:2012uva}
while the radially excited states seem whitin reach
\cite{Aaltonen:2013atp}. Very recently, the CDF Collaboration
studied the orbitally excited $B$ mesons, and reported that the new
$B(5790)$ state could be the radially excited state in the bottom
meson family \cite{Aaltonen:2013atp}. There are several theoretical
studies of the bottom and bottom-strange mesons before
\cite{Falk:1995th,Orsland:1998de} and after
\cite{Zhong:2008kd,Luo:2009wu} the experimental observation of these
states.

Now is a good time to carry out a comprehensive theoretical
study on higher bottom and bottom-strange mesons. In this work, we
will calculate the mass spectrum of higher bottom and bottom-strange
mesons and the corresponding two-body strong decay behavior. We hope
the present investigation may not only shed light on the properties
of the observed bottom and bottom-strange states, but also provide
valuable clues to further experimental exploration of the radially
and orbitally excited bottom and bottom-strange states.

This paper is organized as follows. After the Introduction, we
present the analysis of the mass spectrum of the bottom and
bottom-strange meson family in comparison with the available
experimental data and other theoretical results. In Sec. \ref{sec3},
we discuss the two-body strong decay behavior of the higher bottom
and bottom-strange mesons. The last section is devoted to the
discussion and Conclusion.

\section{The mass spectrum} \label{sec2}

We first calculate the mass spectrum of the higher bottom and
bottom-strange mesons in the framework of the relativistic quark
model \cite{Godfrey:1985xj}, where the total Hamiltonian
$\tilde{H}_1$ describes the interaction between quark and anti-quark
in the meson
\begin{equation}
\tilde{H}_1=\left(p^2+m_1^2\right)^{1/2}+\left(p^2+m_2^2\right)^{1/2}+\tilde{H}_{12}^{\text{conf}}+\tilde{H}_{12}^{\text{so}}
+\tilde{H}_{12}^{\text{hyp}},
\end{equation}
where $\tilde{H}_{12}^{\text{conf}}$ denotes the confinement term
and $\tilde{H}_{12}^{\text{so}} $ is the spin-orbit term which can
be decomposed into the symmetric part $\tilde{H}_{(12)}^{\text{so}}$
and the antisymmetric part $\tilde{H}_{[12]}^{\text{so}} $. In addition,
$\tilde{H}_{12}^{\text{hyp}}$ is the sum of the tensor and contact
terms, i.e.,
\begin{equation}
\tilde{H}_{12}^{hyp}=\tilde{H}^{\text{tensor}}_{12}+\tilde{H}^{\text{c}}_{12}.
\end{equation}
The concrete forms of these terms can be found in Appendix A of
Ref. \cite{Godfrey:1985xj}.

In the bases $|n^{2S+1}L_J\rangle$, the antisymmetric part of the
spin-orbit term  $\tilde{H}_{[12]}^{\text{so}} $ and the tensor term
$\tilde{H}^{\text{tensor}}_{12}$ have nonvanishing off-diagonal
elements, which result in the mixing of the states with quantum
numbers $^{3}L_J$ and $^{1}L_J$  or with $^{3}L_J$ and
$^{3}(L\pm2)_J$. Thus, the total Hamiltonian  $\tilde{H}_1$ can be
divided into two parts in this bases, which include the diagonal
part $H_{\text{diag}}$ and off-diagonal part $H_{\text{off}}$ with
the form
\begin{equation}
H_{\text{off}}=\tilde{H}_{[12]}^{\text{so}}+\left(\tilde{H}^{\text{tensor}}_{12}\right)_{\text{off}},
\end{equation}
where  $\left(\tilde{H}_{12}^{\text{tensor}}\right)_{\text{off}}$
denotes the off-diagonal parts of $\tilde{H}_{12}^{\text{tensor}}$.
In the following, we first diagonalize $H_{\text{diag}}$ in the
simple harmonic oscillator bases and obtain the eigenvalues and
eigenvectors corresponding to the wave function of the meson. One
also needs to diagonalize the off-diagonal part $H_{\text{off}}$ in
the bases $|n^{2S+1}L_J\rangle$, which is treated as the
perturbative term. We neglect the perturbative term $H_{\text{off}}
$ in the present calculation.

The free parameters of the adopted relativistic quark
model  are listed in Table II of Ref. \cite{Godfrey:1985xj}, which include the quark masses
and the coefficients in the effective potential. Here, we list the quark masses
\begin{equation}
   % \nonumber to remove numbering (before each equation)
     m_u =m_d=220~\text{MeV},~
     m_b = 4977~\text{MeV},~
     m_s = 419~\text{MeV},
   \end{equation}
   which are also
applied in the following two-body strong decay calculation.

With the above preparation, we obtain the mass spectra of the bottom
and bottom-strange meson families as shown in Fig. \ref{spectrum},
where the masses of the $1S$, $2S$, $3S$, $1P$, $2P$, $1D$, $2D$,
and $1F$ states are given. Godfrey {\it et al.} calculated the mass
spectrum of some of the bottom and bottom-strange mesons long ago
\cite{Godfrey:1985xj}. We also notice that there exist predictions
of the mass spectra of the bottom and bottom-strange mesons using
other theoretical models. For example, Ebert {\it et al.} adopted
the relativistic quark model based on the quasipotential approach
\cite{Ebert:2009ua} while the authors of Ref. \cite{Di
Pierro:2001uu} used the relativistic quark model with the Dirac
Hamiltonian potential, where the correction of $1/m_{b}$ in the
potential is also included. In addition, in Refs. \cite{Matsuki:2006rz,Matsuki:2011xp} the masses of the radial excitations of $D$/$D_s$/$B$/$B_s$
were predicted, which are comparable with the corresponding results listed in Table \ref{spectrum}.

\begin{figure}[htbp]
\centering%
\begin{tabular}{c}
\scalebox{0.75}{\includegraphics{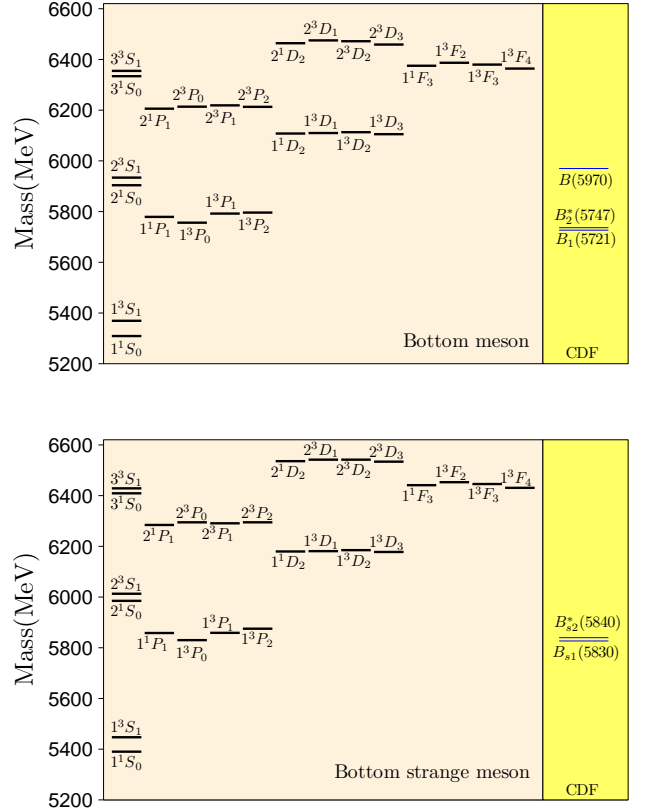}}
\end{tabular}
\caption{(color online). The comparison of the mass spectrum of
bottom and bottom-strange meson calculated by the relativistic quark
model (see Table \ref{spectrum} for the concrete values) with the
experimental data
\cite{Abreu:1994hj,Akers:1994fz,Buskulic:1995mt,Barate:1998cq,Abazov:2007vq,Aaltonen:2008aa,Aaltonen:2007ah,Abazov:2007af,Aaij:2012uva,Aaltonen:2013atp}.}
 \label{mass}
\end{figure}

In Fig. \ref{mass} and Table \ref{spectrum}, we list our results of
the mass spectra of the bottom and bottom-strange meson families,
the results from other groups, and compare them with the
experimental data.

\renewcommand{\arraystretch}{1.3}
\begin{table*}[htbp]\centering\caption{ The obtained masses of the bottom and bottom-strange mesons
in this work in comparison with the experimental values
\cite{Beringer:1900zz} and other theoretical results
\cite{Ebert:2009ua,Di Pierro:2001uu}. In the second and sixth
columns, the values listed in the brackets are the $\beta$ values in
the simple harmonic oscillator (SHO) wave function, which will be
adopted in studying the strong decays of these higher bottom and
bottom-strange mesons. Here, the $\beta$ value in the SHO wave
function is obtained by reproducing the realistic root-mean-square
(rms) radius given by this work. In addition, we want to emphasize
that the results from Refs. \cite{Ebert:2009ua,Di Pierro:2001uu}
consider the mixing of the states with the same $J^P$ quantum
number. }\label{spectrum}
\begin{tabular}{c|cccc|cccc}
\toprule[1pt]
 \multirow{2}{*}{$n^{2S+1}L_J$}
   &\multicolumn{4}{ c |}{Bottom meson}&\multicolumn{4}{ c }{Bottom-strange meson}\\
  % \cline{2-9}
   &This work&Ref. \cite{Ebert:2009ua}&Ref. \cite{Di Pierro:2001uu}&Expt. \cite{Beringer:1900zz}&This work&Ref. \cite{Ebert:2009ua}&Ref. \cite{Di Pierro:2001uu}&Expt. \cite{Beringer:1900zz}\\  \midrule[1pt]
$1^1S_0$&5309(0.63)&5280&5279&$5279.55\pm0.26$&5390(0.69)&5372&5373&$5366.7\pm0.4$\\
$1^3S_1$&5369(0.57)&5326&5324&$5325.2\pm0.4$&5447(0.63)&5414&5421&$5415.8\pm1.5$  \\
$2^1S_0$&5904(0.51)&5890&5886&--&5985(0.54)&5976&5985&--\\
$2^3S_1$&5934(0.50)&5906&5920&--&6013(0.53)&5992&6019   &--\\
$3^1S_0$&6334(0.47)&6379&6320&--&6409(0.49)&6467&6421&--\\
$3^3S_1$&6355(0.46)&6387&6347&--&6429(0.48)&6475&6449 &--\\
$1^3P_2$&5796(0.49)&5741&5714&$5743\pm5$&5875(0.52)&5842&5820&$5839.96\pm0.20$  \\
$1^3P_0$&5756(0.56)&5749&5706&--&5830(0.59)&5833&5804&--\\
$1^3P_1$&5782(0.53)&5723&5700&$5723.5\pm2.0$&5859(0.56)&5831&5805&$5828.7\pm0.4$ \\
$1^1P_1$&5779(0.52)&5774&5742&--&5858(0.55)&5865&5842&--\\
$2^3P_2$&6213(0.46)&6260&6188&--&6295(0.49)&6359&6292&-- \\
$2^3P_0$&6214(0.49)&6221&6163&--&6279(0.51)&6318&6292 &-- \\
$2^3P_1$&6219(0.48)&6281&6175&--&6291(0.50)&6345&6278 &-- \\
$2^1P_1$&6206(0.48)&6209&6194&--&6284(0.50)&6321&6296 &--\\
$1^3D_3$&6105(0.46)&6091&5933&--&6178(0.48)&6191&6103&--\\
$1^3D_1$&6110(0.50)&6119&6025&--&6181(0.52)&6209&6127&--\\
$1^3D_2$&6113(0.48)&6121&5985&--&6185(0.50)&6218&6095&--\\
$1^1D_2$&6108(0.48)&6103&6025&--&6180(0.50)&6189&6140 &--\\
$2^3D_3$&6459(0.44)&6542& --   &--&6534(0.46)&6637&  --   &--\\
$2^3D_1$&6475(0.47)&6534&    --&--&6542(0.48)&6629&    --&--\\
$2^3D_2$&6472(0.46)&6554&   -- &--&6542(0.47)&6651&  -- &--\\
$2^1D_2$&6464(0.46)&6528&   -- &--&6536(0.47)&6625&   --&--\\
$1^3F_4$&6364(0.44)&6380&6226&--&6431(0.45)&6475&6337&-- \\
$1^3F_2$&6387(0.47)&6412&6264&--&6453(0.48)&6501&6369 &--\\
$1^3F_3$&6380(0.45)&6420&6220&--&6446(0.47)&6515&6332 &--\\
$1^1F_3$&6375(0.45)&6391&6271&--&6441(0.47)&6468&6376 &--\\
\bottomrule[1pt]
\end{tabular}
\end{table*}

Until now, there has been only some limited experimental information on
the low-lying bottom and bottom-strange mesons, which are shown in
the fifth and ninth columns. The theoretical results can reproduce
these experimental data well. When comparing our results with those
given by Refs. \cite{Ebert:2009ua,Di Pierro:2001uu}, we notice that
the discrepancy of the values of the mass from different models is
about 50 MeV for the $2S$, $3S$, and $1D$ states, while the mass
difference for the $1F$, $2P$ and  $2D$ states from different model
calculations may reach up to about $100\sim 200$ MeV. Thus, we will
consider the mass dependence of the strong decay width of the higher
bottom and bottom-strange mesons, which will be discussed in the
next section. In the following, we will employ our numerical results
to study the two-body strong decay behavior of these higher bottom
and bottom-strange mesons.

\section{Two-body strong decay behavior}\label{sec3}

In this work, we will study the
two-body strong decay of the higher bottom and bottom-strange
mesons allowed by the Okubo-Zweig-Iizuka (OZI) rule, where the quark pair creation (QPC) model is adopted in our
calculation.

The QPC model \cite{Micu:1968mk,Le
Yaouanc:1972ae,LeYaouanc:1988fx,vanBeveren:1979bd,vanBeveren:1982qb,Bonnaz:2001aj,roberts}
is an effective approach to study the strong decay of hadrons, which
has been widely applied to study the decay behaviors of the newly
observed hadron states
\cite{Zhang:2006yj,Liu:2009fe,Sun:2009tg,Sun:2010pg,Yu:2011ta,Wang:2012wa,Ye:2012gu,He:2013ttg}.
In the QPC model, the meson OZI-allowed decay occurs via the
creation of a quark-antiquark pair from the vacuum. The created
quark-antiquark pair is a flavor and color singlet with the vacuum
quantum number $J^{PC}=0^{++}$. The transition operator $T$ is
written as \cite{Blundell:1996as}
\begin{eqnarray}
 T&=&-3\gamma\sum_{m}\langle1m;1~-m|00\rangle\int d\textbf{k}_3d\textbf{k}_4\delta^3(\textbf{k}_3+\textbf{k}_4) \nonumber \\
&&\times
\mathcal{Y}_{1m}\left(\frac{\textbf{k}_3-\textbf{k}_4}{2}\right)\chi^{34}_{1,-m}
 \phi_0^{34}\omega^{34}_0 d^{\dag}_{3i}(\textbf{k}_3)b^{\dag}_{4j}(\textbf{k}_4),
\end{eqnarray}
where the flavor and color wave functions have the forms
$\phi_0^{34}=(u\bar{u}+d\bar{d}+s\bar{s})/\sqrt{3}$ and
$\omega_0^{34}=\delta_{ij}/\sqrt{3}$, respectively. The color indices are denoted by $i$ and $j$.
The solid harmonic polynomial is $\mathcal{Y}_{lm}(\textbf{k})=|\textbf{k}|^lY_{lm}(\textbf{k})$. The spin wave
function is $\chi^{34}_{1,-m}$ with an angular momentum quantum number $(1,-m)$. $\gamma$
denotes the model parameter, which describes the strength of the
quark-antiquark pair creation from the vacuum. In our calculation,
the $\gamma$ value is chosen to be 6.3 and $6.3/\sqrt{3}$ for the
creations of $u/d$ quark and $s$ quark \cite{Sun:2009tg},
respectively.

The helicity amplitude
$\mathcal{M}^{M_{J_A}M_{J_B}M_{J_C}}$(\textbf{K}) of the $A\to B+C$
decay is defined as
\begin{equation}
\langle
BC|T|A\rangle=\delta^3(\textbf{K}_B+\textbf{K}_C-\textbf{K}_A)\mathcal{M}^{M_{J_A}M_{J_B}M_{J_C}}(\textbf{K}),
\end{equation}
which can be further expressed as the product of the flavor, spin
and spatial matrix elements, where the calculation of the spatial
matrix element is the most crucial part in the whole QPC calculation
(see \cite{Blundell:1996as} for more details). In order to obtain
the spatial matrix element, we adopt the mock state to describe the
mesons \cite{Hayne:1981zy}. In our calculation, the spatial radial
wave function of the meson is represented by a SHO wave function \cite{Blundell:1996as} with a
parameter $\beta$. The $\beta$ values in the SHO wave functions are
determined by reproducing the rms radius of the corresponding states
calculated in the relativistic quark model. In Table \ref{spectrum},
we list these adopted $\beta$ values for the bottom and
bottom-strange mesons discussed.

Converting the helicity amplitude
$\mathcal{M}^{M_{J_A}M_{J_B}M_{J_C}}(\textbf{K})$ to the partial
wave amplitude $\mathcal{M}_{SL}(|\textbf{K}|)$ through the
Jacob-Wick formula \cite{Jacob:1959at}, we get
\begin{eqnarray}
\mathcal{M}_{LS}(\textbf{K})&=&\frac{\sqrt{2L+1}}{2J_A+1}\sum_{M_{J_B},M_{J_C}}\langle L0SM_{J_A}|J_AM_{J_A}\rangle \nonumber \\
 &&\times\langle J_BM_{J_B}J_CM_{J_C}|SM_{J_A}\rangle
\mathcal{M}^{M_{J_A}M_{J_B}M_{J_C}}(\textbf{K}).
\end{eqnarray}
Finally, the decay width relevant to the partial wave amplitude is
\begin{equation}
\Gamma=\pi^2\frac{|\textbf{K}|}{M_A^2}\sum_{LS}\left|\mathcal{M}_{LS}(\textbf{K})\right|^2,
\end{equation}
where $M_A$ is the mass of the initial meson $A$. We take the
experimental mass as input if the corresponding decay involves the
observed state. Otherwise, we take the predicted mass from the
relativistic quark mode in Table \ref{spectrum} when studying the
strong decay of the bottom and bottom-strange mesons.

In the following subsections, we illustrate the OZI-allowed strong
decay behaviors of the bottom and bottom-strange mesons in detail.
We use $B$ and $B_s$ to represent bottom and bottom-strange mesons,
respectively.

\subsection{Bottom meson}

\subsubsection{$1P$ and $2P$ states}
\renewcommand{\arraystretch}{1.3}
\begin{table}[htbp]\centering
\caption{The calculated partial and total decay widths (in units of
MeV) of $2P$ states in the $B$ meson family.  The forbidden decay
channels are marked by  --. For the $2P(1^+)/2P^\prime(1^+)$ states,
we use $\square$ to mark the allowed decay channels. \label{tab:B}}
\begin{tabular}{lccccccc}
\toprule[1pt]
 Channels       &$2^3P_0$      &$2^3P_2$             &$2P(1^+)$/$2P^\prime(1^+)$             \\   \midrule[1pt]
$\pi B$         &47            &$9.7\times 10^{-2}$  &--                                     \\
$\pi B^*$       &--            &1.4                  &$\square$                               \\
$\pi B(1^3P_0)$ &--            &--                   &$\square$                                \\
$\pi B(1^3P_2)$ &--            &4.6                  &$\square$                                \\
$\pi B(1P)$     &10            &16                   &$\square$                              \\
$\pi B(1P^\prime)$&51          &1.3                  &$\square$                             \\
$\eta B$        &5.6           &0.11                 &--                            \\
$\eta B^*$      &--            &0.45                 &$\square$                      \\
$\rho B$        &--            &4.3                  &$\square$                      \\
$\rho B^*$      &27            &26                   &$\square$                          \\
$\omega B$      &--            &1.3                  &$\square$                          \\
$\omega B^*$    &8.8           &9.3                  &$\square$                         \\
$K B_s$         &3.6           &$1.3\times 10^{-2}$  &--                                \\
$K B_s^*$       &--            &0.11                 &$\square$                       \\
$K^* B_s$       &--            &$2.7\times 10^{-2}$  &--
\\ \midrule[1pt] Total            &154          &64
&--         \\  \bottomrule[1pt]
\end{tabular}
\end{table}

Although the predicted mass of the $1^3P_0$ bottom meson is above
both $B\pi$ and $B^*\pi$ thresholds, $B(1^3P_0)$ can only decay into
$B\pi$. Its $B^*\pi$ decay is forbidden due to the parity
conservation. The partial decay width of $B(1^3P_0)\to B\pi$ can
reach up to 225 MeV. This state is very broad, which provides a
natural explanation of why $B(1^3P_0)$ is still missing
experimentally. There were other theoretical calculations of the
decay width of the $B(1^3P_0)$ state. For example, the authors used
the chiral quark model and obtained $\Gamma(B(1^3P_0)\to B\pi)=272$
MeV \cite{Zhong:2008kd}, which is consistent with our result.

The two $J^P=1^+$ states are the mixture of the $1^1P_1$ and
$1^3P_1$ states
\begin{equation}
\left(
  \begin{array}{c}
   1P(1^+) \\
   1P'(1^+)  \\
  \end{array}
\right) = \left(
\begin{array}{cc}
   \cos\theta_{1P}&\sin\theta_{1P} \\
   -\sin\theta_{1P}&\cos\theta_{1P}  \\
\end{array}
\right) \left(
  \begin{array}{c}
   1 ^1P_1 \\
    1^3P_1 \\
  \end{array}
\right),\label{pmix}
\end{equation}
where $\theta_{1P}$ denotes the mixing angle. In general, in the
heavy quark limit one has
$\theta_{1P}=-54.7^\circ$\cite{Close:2005se}. However, the obtained
$\theta_{1P}$ value deviates from this ideal value when the heavy
quark mass is finite in the case of the bottom and bottom-strange
mesons. We will discuss this issue later.

In heavy quark effective theory, the heavy-light meson system can be
categorized into different doublets by the angular momentum of the
light component $j_\ell$, which is a good quantum number in the
limit of $m_Q\to\infty$. There are two $1^+$ states in the
$(0^+,1^+)$ and $(1^+,2^+)$ P-wave doublets, which correspond to
$j_\ell=1/2^+$ and $j_\ell=3/2^+$, respectively. We can easily
distinguish these two $1^+$ states; i.e., the $1^+$ state in the
$(0^+,1^+)$ doublet is broad while the other $1^+$ state in the
$(1^+,2^+)$ doublet is narrow. In Eq. (\ref{pmix}), the $1P(1^+)$
and $1P^\prime(1^+)$ states correspond the $1^+$ states in the
$(0^+,1^+)$ and $(1^+,2^+)$ doublets, respectively. The following
quantitative study of the decay behaviors will confirm the above
qualitative observation.

In 2007, D0 Collaboration reported the observation of $B_1(5721)^0$
in the $B^{*+}\pi^-$ channel, where its measured mass is
$5720.6\pm2.4\pm1.4$ MeV \cite{Abazov:2007vq}. Later, CDF confirmed
$B_1(5721)^0$ with the mass ${5725.3^{+1.6+1.4}_{-2.2-1.5}}$ MeV
\cite{Aaltonen:2008aa}. Very recently, CDF \cite{Aaltonen:2013atp}
studied the orbitally excited $B$ mesons by using the full CDF Run
II data sample. Besides confirming $B_1(5721)$, CDF reported the
width of $B_1(5721)$ for the first time, where the measured width is
$20\pm2\pm5$ MeV and $42\pm11\pm13$ MeV for the neutral and charged
$B_1(5721)$ states \cite{Aaltonen:2013atp}, respectively. Since
$B_1(5721)$ has a narrow width, $B_1(5721)$ is a good candidate of
the $1P^\prime(1^+)$ state in the $B$ meson family.

\begin{figure}[htbp]
\centering%
\begin{tabular}{c}
\scalebox{0.62}{\includegraphics{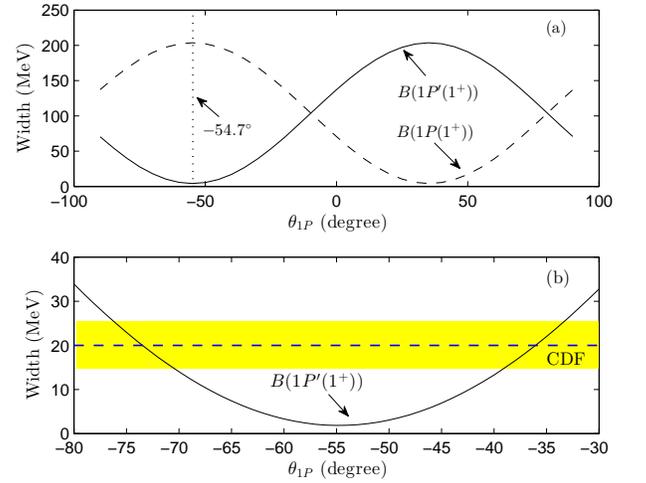}}
\end{tabular}
\caption{(color online). (a) The dependence of the total decay
widths of $B(1P(1^+))$ (dashed curve) and $B(1P^\prime(1^+))$ (solid
curve) on the mixing angle $\theta_{1P}$. (b) The dependence of the
total width of $B(1P^\prime(1^+))$ on the mixing angle
$\theta_{1P}=(-80\sim-30)^\circ$ and the comparison with the CDF
data \cite{Aaltonen:2013atp}. Here, the vertical dashed line in (a) corresponds to the ideal mixing angle
$\theta_{1P}=-54.7^\circ$. }
 \label{B15721}
\end{figure}

For $B_1(5721)$, its dominant decay is $B^*\pi$. With the QPC model,
we calculate $B_1(5721)^0\to B^*\pi$. In Fig. \ref{B15721}, the
dependence of the calculated width of $B_1(5721)^0$ on the mixing
angle $\theta_{1P}$ is given. If taking the ideal value
$\theta_{1P}=-54.7^\circ$, the corresponding width of $B_1(5721)^0$
is 1.8 MeV, which is far smaller than the experimental data. We find
that the calculated width of $B_1(5721)^0$ overlaps with the
experimental value when $\theta_{1P}$ is in the range of
$(-77^\circ\sim-70^\circ)$ or ($-40^\circ\sim -33^\circ$). We also
predict the width of $B(1P(1^+))$ state, where is
$\Gamma(B(1P(1^+)))=170\sim 197$ MeV corresponding to
$\theta_{1P}=(-77^\circ\sim-70^\circ)$ or ($-40^\circ\sim
-33^\circ$), which confirms $B(1P(1^+))$ is a broad state.

$B_2^*(5747)$ was first observed by D0 in its $B^{*+}\pi^-$ and
$B^+\pi^-$ channels \cite{Abazov:2007vq}. Other experimental
information of $B_2^*(5747)^0$ given by D0 includes:
$M(B_2^*(5747)^0)=5746.8\pm2.4\pm1.7$ MeV and
$\Gamma(B_2^*(5747)^0\to B^{*+}\pi^-)/\Gamma(B_2^*(5747)^0\to
B^{(*)+}\pi^-)=0.475\pm0.095\pm0.069$ \cite{Abazov:2007vq}. In 2009,
CDF announced the confirmation of $B_2^*(5747)^0$ with mass
$M=5740.2^{+1.7+0.9}_{-1.8-0.8}$ MeV and width
$\Gamma=22.7^{+3.8+3.2}_{-3.2-10.2}$ MeV  \cite{Aaltonen:2008aa}. In
Ref. \cite{Aaltonen:2013atp}, CDF again carried out the study of
$B_2^*(5747)$. The mass and width are
$M=5736.6\pm1.2\pm1.2\pm0.2/5737.1\pm1.1\pm0.9\pm0.2$ MeV and
$\Gamma=26\pm3\pm3/17\pm6\pm8$ MeV, respectively for the
neutral/charged $B_2^*(5747)$ states \cite{Aaltonen:2013atp}.

As a $J^P=2^+$  $B$ meson, $B_2^*(5747)$ decays into the $B\pi$ and
$B^*\pi$ channels. In the QPC model, the total decay width of
$B_2^*(5747)^0$ is around 3.7 MeV, which is composed of
$\Gamma(B_2^*(5747)^0\to B\pi)=1.9$ MeV and $\Gamma(B_2^*(5747)^0\to
B^*\pi)=1.8$ MeV. We notice that our result of the total decay width
of $B_2^*(5747)^0$ is smaller than the experimental central value
\cite{Aaltonen:2008aa,Aaltonen:2013atp}. As shown in Refs.
\cite{Aaltonen:2008aa,Aaltonen:2013atp}, there exist large
experimental errors for the measured width of $B_2^*(5747)$. Thus,
further experimental measurement of the resonance parameter of
$B_2^*(5747)$ will be helpful to clarify this difference. In
addition, we also obtain the ratio
\begin{equation}
\frac{B(B_2^*(5747)\rightarrow
B^{*+}\pi^-)}{B(B_2^*(5747)\rightarrow B^{+}\pi^-)}=0.95,
\end{equation}
which is consistent with the experimental value $1.10\pm
0.42\pm0.31$ \cite{Abazov:2007vq}.

For the $2P$ states in the $B$ meson family, more decay channels are
open. However, there has been no experimental observation up to now. We
list our predictions of the partial and total decay widths in Table
\ref{tab:B}.

For the $B(2^3P_0)$ meson, the dominant decay modes are $\pi
B(1P^\prime)$, $\pi B$, and $\rho B^*$. In addition, $B(2^3P_0)$ can
decay into $\omega B^*$, $KB_s$, $\eta B$, and $\pi B(1P)$. The sum
of all partial decay widths of $B(2^3P_0)$ reaches up to 154 MeV,
which indicates $B(2^3P_0)$ is a broad state. Among the dominant
decay channels of $B(2^3P_0)$, $\pi B$ is the most suitable channel
for searching $B(2^3P_0)$.

$B(2^3P_2)$ mainly decays into $\rho B^*$, $\pi B(1P)$, $\omega
B^*$, and $\pi B(1^3P_2)$, where the partial width of $B(2^3P_2)\to
\rho B^*$ is the largest since $B(2^3P_2)\to \rho B^*$ occurs via
S-wave. In contrast, the D-wave decay modes $B(2^3P_2)\to \pi
B,\,KB_s,\,K^*B_s$ can be neglected. The width of $B(2^3P_2)$ is
$64$ MeV.

As the mixture of the $2^1P_1$ and $2^3P_1$ states, the two $1^+$
states $B(2P(1^+))$ and $B(2P^\prime(1^+))$ satisfy the following
relation
\begin{equation}
\left(
  \begin{array}{c}
   2P(1^+) \\
   2P'(1^+)  \\
  \end{array}
\right) = \left(
\begin{array}{cc}
   \cos\theta_{2P}&\sin\theta_{2P} \\
   -\sin\theta_{2P}&\cos\theta_{2P}  \\
\end{array}
\right) \left(
  \begin{array}{c}
   2^1P_1 \\
    2^3P_1 \\
  \end{array}
\right),\label{2pmix}
\end{equation}
where $\theta_{2P}$ denotes the mixing angle. The allowed decay
modes of $B(2P(1^+))$ and $B(2P^\prime(1^+))$ are shown in Table
\ref{tab:B}.

\begin{figure}[htbp]
\centering%
\begin{tabular}{c}
\scalebox{0.59}{\includegraphics{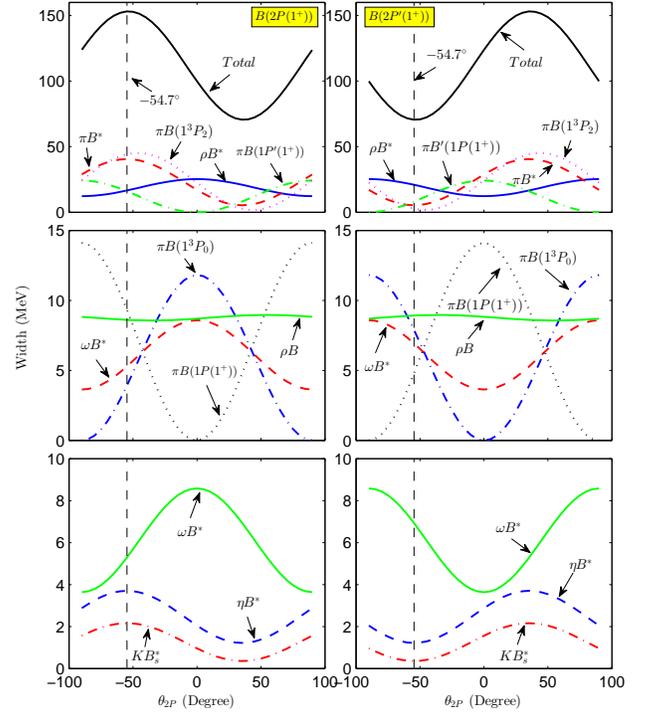}}
\end{tabular}
\caption{(color online). The dependence of the decay behavior of
$B(2P(1^+))$ (the first column) and $B(2P^\prime(1^+))$ (the second
column) on the mixing angle $\theta_{2P}$.}
 \label{b2p}
\end{figure}

We obtain the total and partial decay widths of $B(2P(1^+))$ and
$B(2P^\prime(1^+))$, which depend on the mixing angle $\theta_{2P}$
(see Fig. \ref{b2p} for details). Since the experimental
information of $B(2P(1^+))$ and $B(2P^\prime(1^+))$ is absent, in
the following discussion we take the typical value
$\theta_{2P}=-54.7^\circ$. We find that the total decay width of
$B(2P(1^+))$ ($\Gamma(B(2P(1^+)))=153$ MeV) is larger than that of
$B(2P^\prime(1^+))$ ($\Gamma(B(2P^\prime(1^+)))=70$ MeV), which is
consistent with the estimate under the heavy quark effective theory.
For $B(2P(1^+))$ and $B(2P^\prime(1^+))$, its main decay channels
include $\pi B^*$, $\pi B(1^3P_2)$, $\pi B(1P^\prime(1^+))$, $\rho
B^*$ and $\pi  B(1P(1^+))$. In addition, the other decay modes of
$B(2P(1^+))$ and $B(2P^\prime(1^+))$ are also presented in Fig.
\ref{b2p}.

\subsubsection{$2S$ and $3S$ states}

Very recently, the CDF Collaboration reported the evidence of a new
resonance $B(5970)$ in analyzing both $B^0\pi^+$ and $B^+\pi^-$ mass
distributions \cite{Aaltonen:2013atp}. The mass and width of
$B(5970)$ are
\begin{eqnarray}
(M, \Gamma)_{B(5970)^0}& =&(5978\pm5\pm12,\, 70\pm18\pm31 )\,\text{MeV}, \nonumber \\
(M,\Gamma)_{B(5970)^+}&=&(5961\pm5\pm12,
\,60\pm20\pm40)\,\text{MeV}, \nonumber
\end{eqnarray}
which correspond to the neutral and charged $B(5970)$, respectively.
The comparison of the mass of $B(5970)$ and the mass spectrum in
Table \ref{spectrum} indicates that the mass of $B(5970)$ is close
to the estimated masses of the $2^1S_0$ and $2^3S_1$ states of the
$B$ meson family. Since $B(5970)$ can decay into $B\pi$, we can
exclude the $B(2^1S_0)$ assignment of $B(5970)$.

\renewcommand{\arraystretch}{1.3}
\begin{table}[htbp]\centering
\caption{The partial and total decay widths (in units of MeV) of
$2S$ and $3S$ states in the $B$ meson family. Here, we adopt -- to
denote the forbidden decay channels. }\label{1s2s}
\begin{tabular}{lcccccc}
\toprule[1pt] Channels                 &$2^1S_0$
&$2^3S_1$              &$3^1S_0$           &$3^3S_1$
\\  \midrule[1pt]
$\pi B$                  &--                  &9.1                   &--                 &2.6                           \\
$\pi B^*$                &33                  &23                    &8.8                &6.1                           \\
$\pi B(1^3P_0)$          &3.9                 &--                    &3.7                &--                           \\
$\pi B(1^3P_2)$          &$2.0\times 10^{-3}$  &$1.0\times 10^{-2}$   &11.2               &4.9                         \\
$\pi B(1P(1^+))$              &--                  &10                    &--                 &3.9                          \\
$\pi B(1P^\prime(1^+))$       &--                  &0.2                   &--                 &2.8                         \\
$\eta B$                 & --                 &1.2                   &--                 &0.49                          \\
$\eta B^*$               &0.62                &2.0                   &1.0                &0.87                        \\
$\eta B(1^3P_0)$         &--                  &--                    &0.16               &--                       \\
$\eta B(1^3P_2)$         &--                  &--                    &$7.2\times 10^{-2}$&$8.8\times 10^{-2}$      \\
$\eta B(1P(1^+))$             &--                  &--                    &--                 &0.28                     \\
$\eta B(1P^\prime(1^+)) $     &--                  &--                    &--                 &0.11                     \\
$\rho B$                 &--                  &--                    &1.3                &1.3                       \\
$\rho B^*$               &--                  &--                    &9.4                &7.0                        \\
$\omega B$               &--                  &--                    &0.49               &$5.7\times 10^{-2}$         \\
$\omega B^*$             &--                  &--                    &3.3                &1.7                        \\
$K B_s$                  &--                  & 0.43                 &--                 &0.24                               \\
$K B_s^*$                &--                  & 0.58                 &0.45               &0.41                         \\
$K^* B_s$                &--                  &--                    &0.46               &0.27                        \\
$K^* B_s^*$              &--                  &--                    &0.49               &0.90                     \\
$KB_s(1^3P_0)$           &--                  &--                    &$4.0\times 10^{-2}$&--                      \\
$K B_s(1^3P_2)$          &--                  &--                    &--                 &$2.8\times 10^{-3}$     \\
$K B_s(1P(1^+))$              &--                  &--                    &--                 &0.10                    \\
$K B_s(1P^\prime(1^+))$       &         --           &--
&--                 &$3.7\times 10^{-3}$      \\  \midrule[1pt]
Total                    &38                  & 47                   & 41                &33                            \\
\bottomrule[1pt]
\end{tabular}
\end{table}

The obtained total width of $B(2^3S_1)$ is 47 MeV, which is in
agreement with the experimental width of $B(5970)$ if one considers
the experimental error. The calculation of the partial decay widths
of $B(2^3S_1)$ indicates that $\pi B^*$, $\pi B$ and $\pi
B(1P(1^+))$ are its main decay channels. The results of the other
partial decay widths of $B(2^3S_1)$ are listed in Table \ref{1s2s}.
At present, $B(5970)$ was only observed in the $\pi B$ channel. Our
study indicates that the $B(5970)$ is very probably $B(2^3S_1)$. We
also suggest the experimental search for $B(5970)$ via its $\pi B^*$
decay.

\renewcommand{\arraystretch}{1.3}
\begin{table}[htbp]\centering
\caption{The partial and total decay widths (in units of MeV) of
$1D$ and $2D$ states in the $B$ meson family. The forbidden decay
channels are marked by  --. For the $1D(2^-)/1D^\prime(2^-)$ and
$2D(2^-)/2D^\prime(2^-)$ states, we use $\square$ to mark the
allowed decay channels. Here, the value $a \times 10^{-b}$ is
abbreviated as $a[b]$.}\label{1d2d}
\begin{tabular}{lcccccccccccc}
\toprule[1pt]
\multirow{2}{*}{Channels}     &\multirow{2}{*}{$1^3D_1$}     &  \multirow{2}{*}{$1^3D_3$}   &\multirow{2}{*}{ $2^3D_1$}           &\multirow{2}{*}{$2^3D_3$}           &$1D(2^-)$  &$2D(2^-)$    \\
&&&&  &  $1D'(2^-)$  &$2D'(2^-)$     \\   \midrule[1pt]
$\pi B$                 &69                   &4.9                  &27                  &1.2                &--         &--              \\
$\pi B^*$               &34                   &6.2                  &12                  &0.49               &$\square$  &$\square$      \\
$\pi B(1^3P_0)$         &--                   &--                   & --                 & --                &$\square$  &$\square$      \\
$\pi B(1^3P_2)$         &1.6                  &0.74                 &5.3                 &0.21               &$\square$  &$\square$     \\
$\pi B(1P(1^+))$             &$6.8[2]$            &$9.0[2]$            &8.1                 &6.4                &$\square$  &$\square$      \\
$\pi B(1P^\prime(1^+))$      &147                  &0.17                 &28                  &$7.5[2]$          &$\square$  &$\square$      \\
$\eta B$                &12                   &0.21                 &4.3                 &$5.8[2]          $&--         &--               \\
$\eta B^*$              &5.6                  &0.20                 &1.6                 &$1.2[3]$          &$\square$  &$\square$        \\
$\eta B(1^3P_0)$        &--                   &--                   & --                 & --                &--         &$\square$     \\
$\eta B(1^3P_2)$        &--                   &--                   &0.62                &0.19               &--         &$\square$       \\
$\eta B(1P(1^+))$            &--                   &--                   &0.74                &0.57               &--         & $\square$     \\
$\eta B(1P^\prime(1^+))$     &--                   &--                   &0.65                &$2.1[2]$&--         &$\square$    \\
$\rho B$                &8.3                  &$1.8[2]$            &0.12               &2.0                 &$\square$  &$\square$     \\
$\rho B^*$              &0.93                 &1.3                  &23                 &6.1                 &$\square$  &$\square$         \\
$\omega B$              &2.4                  &$3.7[3]$            &$3.2[2]$          &0.67                &$\square$  &$\square$       \\
$\omega B^*$            &0.10                 &--                   &7.5                &2.0                 &$\square$  &$\square$    \\
$K B_s$                 &7.4                  & 5.4[2]             &2.5                &$4.4[2]$           &--         &--             \\
$K B_s^*$               &3.3                  &$4.5[2]$            &0.86               &$8.6[3]$           &$\square$  &$\square$      \\
$K^* B_s$               &--                  &--                   &0.25               &$1.9[3]$           &--         &$\square$        \\
$K^* B_s^*$             &--                   &--                   &0.65               &0.75                &--         &$\square$       \\
$KB_s(1^3P_0)$          &--                   &--                   &--                 & --                 &--         &$\square$         \\
$K B_s(1^3P_2)$         &--                   &--                   &0.20               &$4.5[2]$           &--         &$\square$          \\
$K B_s(1P(1^+))$             &--                   &--                   &0.19               &0.12                &--         &$\square$         \\
$K B_s(1P^\prime(1^+))$      &--                   &--
&0.14               &$4.2[3]$           &--         &$\square$   \\
\midrule[1pt]
Total                   &294                  &14                   &122                &23                  &--         &--              \\
\bottomrule[1pt]
\end{tabular}
\end{table}

As the partner of $B(2^3S_1)$, $B(2^1S_0)$ dominantly decays into
$\pi B^*$. In addition, there exists a considerable partial width of
$B(2^1S_0)\to \pi B(1^3P_0)$. Compared with the $\pi B^*$ and $\pi
B(1^3P_0)$, the remaining two decay channels $\eta B^*$ and $\pi
B(1^3P_2)$ can be ignored. Thus, $\pi B^*$ is the ideal channel to
search for $B(2^1S_0)$ in future experiments.

The masses of the two $3S$ states $B(3^1S_0)$ and $B(3^3S_1)$ are
6334 MeV and 6355 MeV, respectively. As shown in Table \ref{1s2s},
more decay channels are open for $B(3^1S_0)$ and $B(3^3S_1)$. The
main decay modes of $B(3^1S_0)$ are $\pi B(1^3P_2)$, $\rho B^*$,
$\pi B^*$, $\omega B^*$. For $B(3^3S_1)$,
mainly $\rho B^*$, $\pi B^*$,
$\pi B(1^3P_2)$, $\pi B(1P(1^+))$, $\pi B(1P^\prime(1^+))$, $\pi B$,
$\omega B^*$, and $\rho B$ contribute to the total decay width. The partial widths of the modes $\eta B(1^3P_2)$, $\omega
B$, $KB_s(1^3P_2)$ and $KB_s(1P^\prime(1^+))$ are quite small.

\subsubsection{$1D$ and $2D$ states}

The decay behaviors of $1D$ and $2D$ states are given in Table
\ref{1d2d}, where we first list the allowed decay channels for the
$1D(2^-)/1D^\prime(2^-)$ $2D(2^-)/2D^\prime(2^-)$ states.

\begin{figure}[htbp]
\centering%
\begin{tabular}{c}
\scalebox{0.57}{\includegraphics{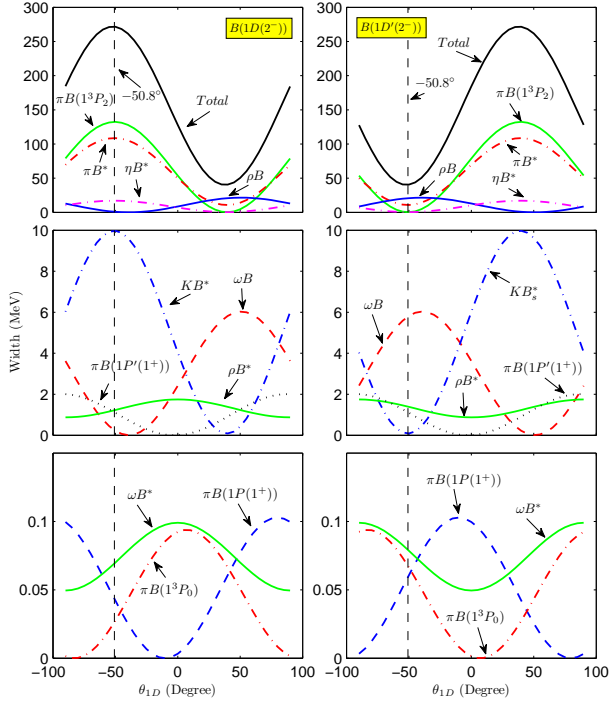}}
\end{tabular}
\caption{(color online). The dependence of the decay behavior of
$B(1D(2^-))$ (the first column) and $B(1D^\prime(2^-))$ (the second
column) on the mixing angle $\theta_{1D}$.}
 \label{1dmixing}
\end{figure}

\begin{figure}[htbp]
\centering%
\begin{tabular}{c}
\scalebox{0.57}{\includegraphics{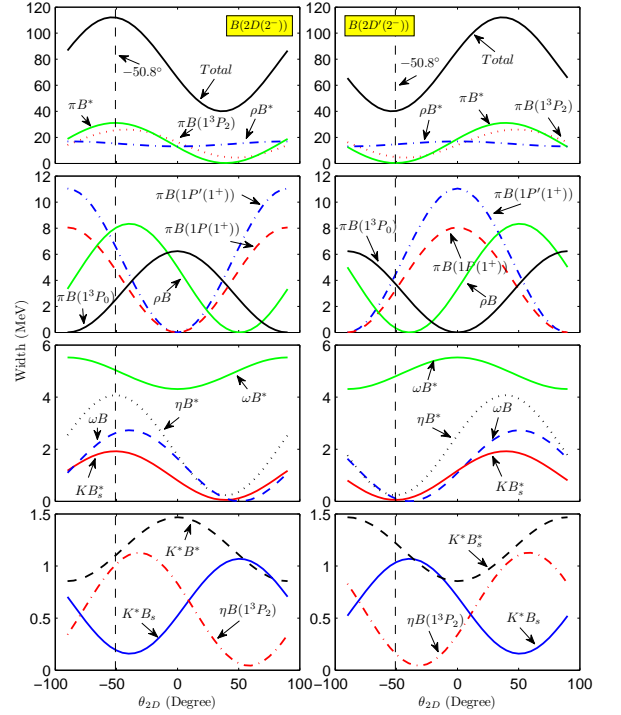}}
\end{tabular}
\caption{(color online). The dependence of the partial and total
decay widths of $B(2D(2^-))$ (the first column) and
$B(2D^\prime(2^-))$ (the second column) on the mixing angle
$\theta_{2D}$.}
 \label{2dmixing}
\end{figure}

$B(1^3D_1)$ is very broad and its total width reaches up to 294 MeV.
$B(1^3D_1)$ dominantly decays into $\pi B(1P^\prime(1^+))$, $\pi B$
and $\pi B^*$. These channels are suitable to search for $B(1^3D_1)$
in future experiments. $B(2^3D_1)$ is the first radial excitation of
$B(1^3D_1)$. We notice that $\pi B(1P^\prime(1^+))$, $\pi B$, $\rho
B^*$, and $\pi B^*$ are the main decay modes of $B(2^3D_1)$, which
renders $B(2^3D_1)$ broad. Different from the broad $B(1^3D_1)$ and
$B(2^3D_1)$, $B(1^3D_3)$ and $B(2^3D_3)$ are rather narrow states
since the total decay widths of $B(1^3D_3)$ and $B(2^3D_3)$ are $14$
and 23 MeV, respectively. For $B(1^3D_3)$, there exist two
dominant decay channels, $\pi B$ and $\pi B^*$. Although more decay
modes are included for $B(2^3D_3)$, its total decay width is not
obviously enhanced. The $\pi B(1P(1^+))$, $\rho B^*$, $\rho B$,
$\omega B^*$ and $\pi B$ are the main decay modes of $B(2^3D_3)$.
The concrete information of the decay pattern of $B(1^3D_1)$,
$B(2^3D_1)$, $B(1^3D_3)$ and $B(2^3D_3)$ can be found in Table
\ref{1d2d}.

$B(nD(2^-))$ and $B(nD^\prime(2^-))$ are the mixture of the $n^3D_2$
and $n^1D_2$ states, which satisfy the following relation:
\begin{equation}
\left(
  \begin{array}{c}
   n D(2^-) \\
    nD^\prime(2^-)  \\
  \end{array}
\right) = \left(
\begin{array}{cc}
    \cos\theta_{nD}&\sin\theta_{nD} \\
    -\sin\theta_{nD}&\cos\theta_{nD}  \\
\end{array}
\right) \left(
  \begin{array}{c}
   n ^1D_2 \\
    n^3D_2\\
  \end{array}
\right)\label{dmix}
\end{equation}
with $n=1,2$.

In Fig. \ref{1dmixing}, we present the dependence of the partial and
total decay widths of $B(1D(2^-))$ and $B(1D^\prime(2^-))$ on the
mixing angle $\theta_{1D}$. In the heavy quark limit, the mixing
angle $\theta_{1D}=-50.8^\circ$  was given in Ref.
\cite{Close:2005se}. $\pi B(1^3P_2)$ and $\pi B^*$ are the dominant
decays for $B(1D(2^-))$ while $\rho B$, $\pi B^*$, and $\omega B$ are
the main contribution to the width of $B(1D^\prime(2^-))$. The QPC
calculation further indicates that the total width of $B(1D(2^-))$
is quite broad. Compared with the total decay width of
$B(1D(2^-))$, the total decay width of $B(1D^\prime(2^-))$ is
narrow. Here, we need to specify that the above conclusion is
obtained with the typical value $\theta_{1D}=-50.8^\circ$. The decay
behaviors of the partial decay widths of $B(2D(2^-))$ and
$B(2D^\prime(2^-))$ depend on the concrete value of $\theta_{1D}$
(see Fig. \ref{1dmixing} for more details).

The variation of the total and partial decay widths of $B(2D(2^-))$
and $B(2D^\prime(2^-))$ with $\theta_{2D}$ is shown in Fig.
\ref{2dmixing}, from which we can obtain the main and subordinate
decays of these two states. The total widths of $B(2D(2^-))$ and
$B(2D^\prime(2^-))$ are 112 and 40 MeV, respectively,
corresponding to $\theta_{2D}=-50.8^\circ$.

\subsubsection{$1F$ states}

There are four $1F$ states. Their decay pattern is listed in Table
\ref{b1f}. We notice that the total decay width of $B(1^3F_2)$ is
quite broad, and can reach up to 254 MeV. Among all decays of
$B(1^3F_2)$, $\pi B(1P^\prime(1^+))$ is the most important channel.
The other main decay modes of $B(1^3F_2)$ are $\pi B$, $\pi B^*$,
$\pi B(1^3P_2)$, $\eta B(1P^\prime(1^+))$, $\rho B$, and $\rho B^*$.
For $B(1^3F_4)$, its total width is estimated as 103 MeV, where
$\rho B^*$ and $\omega B^*$ are its dominant decay mode.

\renewcommand{\arraystretch}{1.2}
\begin{table}[htbp]\centering
\caption{The partial and total decay widths (in units of MeV) of
$1F$ states in $B$ meson family. The forbidden decay channels are
marked by  --. For $1F(3^+)/1F^\prime(3^+)$  states, we use
$\square$ to mark the allowed decay channels. }\label{b1f}
\begin{tabular}{lcccccccccccc}
\toprule[1pt] Channels                   &$1^3F_2$
&$1^3F_4$            &$1F(3^+)$/$1F^\prime(3^+)$  \\
\midrule[1pt]
$\pi B$                    &30                 &3.7                 &--\\
$\pi B^*$                  &21                 &5.3                 &$\square$\\
$\pi B(1^3P_0)$            & --                & --                 &$\square$\\
$\pi B(1^3P_2)$            &14                 &2.5                 &$\square$\\
$\pi B(1P(1^+))$                &0.5                &1.2                 &$\square$\\
$\pi B(1P^\prime(1^+))$         &121                &0.48                &$\square$\\
$\eta B$                   &5.5                &0.31                &--\\
$\eta B^*$                 &3.9                &0.41                &$\square$\\
$\eta B(1^3P_0)$           & --                & --                 &$\square$\\
$\eta B(1^3P_2)$           &0.10               &$4.9\times 10^{-3}$ &--\\
$\eta B(1P(1^+))$               &$1.3\times 10^{-3}$&$7.4\times 10^{-3}$ &$\square$\\
$\eta B(1P^\prime(1^+))$        &11                 &$2.4\times 10^{-3}$ &$\square$\\
$\rho B$                   &15                 &1.1                 &$\square$ \\
$\rho B^*$                 &15                 &56                  &$\square$\\
$\omega B$                 &4.7                &0.33                &$\square$\\
$\omega B^*$               &2.2                &31                  &$\square$\\
$K B_s$                    &2.8                &$7.2\times 10^{-2}$ &--\\
$K B_s^*$                  &2.0                &$9.2\times 10^{-2}$ &$\square$\\
$K^* B_s$                  &0.75               &$1.3\times 10^{-2}$ &$\square$\\
$K^* B_s^*$                &$8.8\times 10^{-2}$&0.24                &--\\
$KB_s(1^3P_0)$             & --                & --                 &$\square$\\
$K B_s(1^3P_2)$            &$1.1\times 10^{-3}$&$1.2\times 10^{-4}$ &--\\
$K B_s(1P(1^+))$                &$5.4\times 10^{-4}$&$2.0\times 10^{-4}$ &$\square$\\
$K B_s(1P^\prime(1^+))$         &3.3                &$4.9\times
10^{-5}$ &$\square$\\ \midrule[1pt]
Total                      &254                &103                 &--\\
\bottomrule[1pt]
\end{tabular}
\end{table}

\begin{figure}[htbp]
\centering%
\begin{tabular}{c}
\scalebox{0.57}{\includegraphics{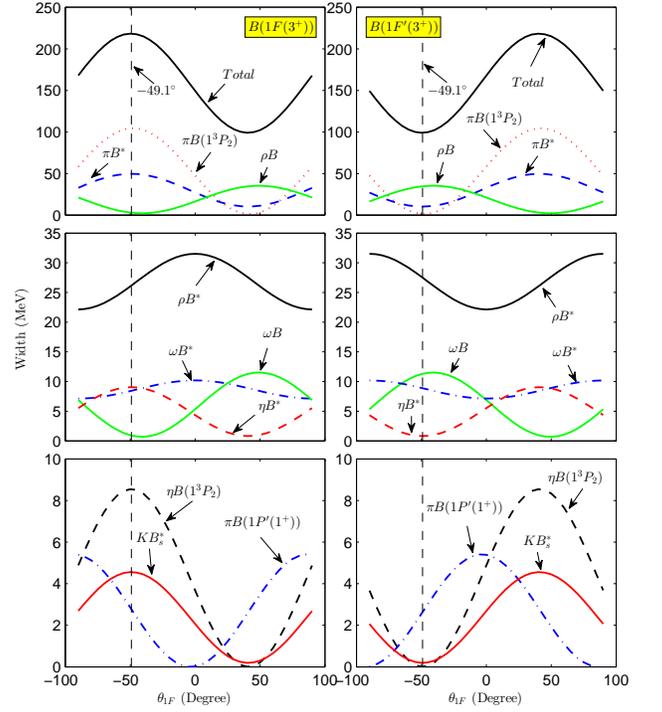}}
\end{tabular}
\caption{(color online). The dependence of the partial and total
decay widths of $B(1F(3^+))$ (the first column) and
$B(1F^\prime(3^+))$ (the second column) on the mixing angle
$\theta_{1F}$.}
 \label{1fmixing}
\end{figure}

In the following, we focus on $B(1F(3^+))$ and $B(1F^\prime(3^+))$,
which are the mixing of the $1^3F_3$ and $1^1F_3$ states,
\begin{equation}
\left(
  \begin{array}{c}
    1F(3^+) \\
    1F^\prime(3^+)  \\
  \end{array}
\right) = \left(
\begin{array}{cc}
    \cos\theta_{1F}&\sin\theta_{1F} \\
    -\sin\theta_{1F}&\cos\theta_{1F}  \\
\end{array}
\right) \left(
  \begin{array}{c}
    1^1F_3 \\
    1^3F_3\\
  \end{array}
\right)\label{fmix}.
\end{equation}
with $\theta_{1F}$=$-49.1^\circ$ determined in heavy quark limit
\cite{Ebert:2009ua}. In Fig. \ref{1fmixing}, the $\theta_{1F}$
dependence of the partial and total decay widths of $B(1F(3^+))$ and
$B(1F^\prime(3^+))$ is presented. If we take the typical value
$\theta_{1F}$=$-49.1^\circ$, we notice that $B(1F(3^+))$ is a quite
broad resonance. From Fig. \ref{1fmixing}, we can easily distinguish
the main decay mode among all allowed decay channels, and we can distinguish which
channel is valuable to further experimental search of the $1F$ state
in the $B$ meson family.

\subsection{Bottom-strange meson}

\subsubsection{$1P$ and $2P$ states}

\renewcommand{\arraystretch}{1.7}
\begin{table*}[htbp]\centering
\caption{The summary of experimental information of
$B_{s1}(5830)$ and $B_{s2}^*(5840)$.}\label{exp-Bs}
\begin{tabular}{cccccc}
\toprule[1pt] State&Collaboration&Mass & Width &Observed decays&
$\frac{\Gamma(B_{s2}^*\rightarrow
B^{*+}K^-)}{\Gamma(B_{s2}^*\rightarrow B^{+}K^-)}$\\ \toprule[1pt]
 \multirow{3}{*}{$B_{s1}(5830)$}
&CDF \cite{Aaltonen:2007ah}  &5829.4 $\pm$ 0.7 MeV           &--      &   $B^{*+}K^-$     &  --\\
&LHCb \cite{Aaij:2012uva} &$5828.40 \pm 0.04 \pm 0.04\pm 0.41$&--  &   $B^{*+}K^-$      &   --    \\
&CDF \cite{Aaltonen:2013atp}&$5828.3 \pm 0.1 \pm 0.1\pm 0.4$&$0.7
\pm 0.3 \pm 0.3$& $B^{*+}K^-$ &--\\ \midrule[1pt]
 \multirow{4}{*}{$B_{s2}^*(5840)$}
 &CDF \cite{Aaltonen:2007ah}  & 5839.6 $\pm$ 0.7&--   &     $B^{+}K^-$          &  --\\
 &D0 \cite{Abazov:2007af}  &5839.6 $\pm$ 1.1 $\pm$ 0.7&--      &      $B^{+}K^-$         &--\\
 &LHCb \cite{Aaij:2012uva}&$5839.99 \pm 0.05 \pm 0.11\pm 0.17$&$1.56 \pm 0.13 \pm0.47$&$B^{*+}K^-$, $B^+K^-$&$0.093 \pm 0.013\pm 0.012$\\
&CDF \cite{Aaltonen:2013atp}&$5839.7 \pm 0.1 \pm 0.1\pm 0.2$ &$2.0 \pm 0.4 \pm 0.2 $& $B^{*+}K^-$,  $B^+K^-$&$0.11\pm 0.03$\\
\bottomrule[1pt]
\end{tabular}
\end{table*}

The decay modes of $B_s(1^3P_0)$ is $BK$. The partial
width of $B_s(1^3P_0)\to BK$ obtained from the QPC model is 225 MeV, which is
almost the same as that from the chiral quark
model in Ref. \cite{Zhong:2008kd} ($\Gamma=227$ MeV). $B_s(1^3P_0)$ is also a broad
state, very similar to $B(1^3P_0)$. Generally speaking, it is
difficult to detect a very broad state experimentally.

The observed $B_{s1}(5830)$ and $B_{s2}^*(5840)$ are good candidates
of $B_s(1P^\prime(1^+))$ and $B_s(1^3P_2)$, respectively. The
experimental information on $B_{s1}(5830)$ and $B_{s2}^*(5840)$ from
the CDF, D0, and LHCb Collaborations
\cite{Aaltonen:2007ah,Abazov:2007af,Aaij:2012uva,Aaltonen:2013atp}
is collected in Table \ref{exp-Bs}.

\begin{figure}[htbp]
\centering%
\begin{tabular}{c}
\scalebox{0.62}{\includegraphics{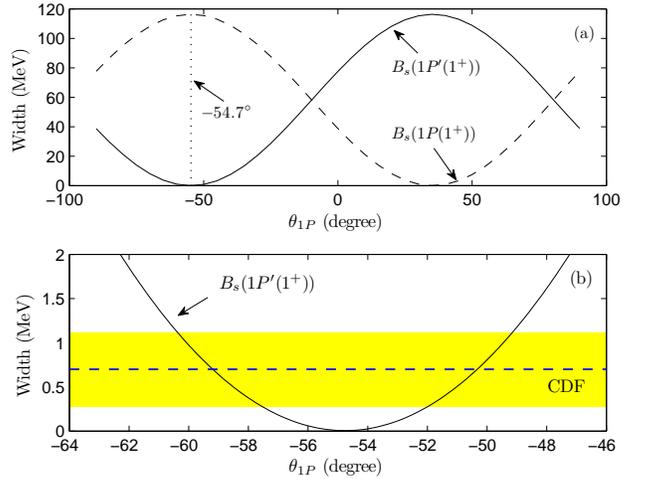}}
\end{tabular}
\caption{(color online). (a) The dependence of the total decay
widths of $B_s(1P(1^+))$ (dashed curve) and $B_s(1P^\prime(1^+))$
(solid curve) on the mixing angle $\theta_{1P}$. (b) The variation
of the total width of $B_s(1P^\prime(1^+))$ with the mixing angle
$\theta_{1P}=(-64\sim-46)^\circ$ and the comparison with the CDF
data \cite{Aaltonen:2013atp}. Here, the vertical dashed line in (a) corresponds to the ideal mixing angle
$\theta_{1P}=-54.7^\circ$. }
 \label{Bs15721}
\end{figure}

If we take the mass of $B_{s2}^*(5840)$ as input, $B_s(1^3P_2)$ can
decay into $BK$ and $B^*K$. With the QPC model, the total decay
width of $B_s(1^3P_2)$ is around $\Gamma(B_s(1^3P_2))=0.26$ MeV and
slightly smaller than the experimental central value which has a
large error \cite{Aaij:2012uva,Aaltonen:2013atp}. However, the
obtained ratio
\begin{eqnarray}
\frac{\Gamma(B_{s2}^*\rightarrow
B^{*+}K^-)}{\Gamma(B_{s2}^*\rightarrow B^{+}K^-)}=0.088
\end{eqnarray}
is in good agreement with the experimental measurement given by LHCb
\cite{Aaij:2012uva} and CDF \cite{Aaltonen:2013atp}, as shown in
Table \ref{exp-Bs}.

In analogy to $B(1P(1^+))$ and $B(1P^\prime(1^+))$, $B_s(1P(1^+))$ and
$B_s(1P^\prime(1^+))$ are the mixture of the $1^3P_1$ and $1^1P_1$
states in the $B_s$ meson family, which also satisfy Eq.
(\ref{pmix}). There only exists the $B^*K$ decay channel for
$B_s(1P(1^+))$ and $B_s(1P^\prime(1^+))$. Thus, we give the dependence
of the total decay widths of $B_s(1P(1^+))$ and $B_s(1P^\prime(1^+))$
on the mixing angle $\theta_{1P}$ (see Fig. \ref{Bs15721}). When the
mixing angle is taken as $\theta_{1P}=-60.5^\circ \sim -57.5^\circ$
or $-52.0^\circ\sim -49.0^\circ$, the total decay width of
$B_s(1P^\prime(1^+)$ overlaps the observed width of $B_{s1}(5830)$
given by CDF \cite{Aaltonen:2013atp}. We notice that there exists a
difference between the realistic and ideal value of the mixing angle
$\theta_{1P}$, which is similar to the situation of $B_1(5721)$. If
adopting $\theta_{1P}=-60.5^\circ \sim -57.5^\circ$ or
$-52.0^\circ\sim -49.0^\circ$, the total decay width of
$B_s(1P(1^+)$ is about 110 MeV. Hence, $B_s(1P(1^+)$ is a broad
resonance, consistent with the rough estimate in heavy quark limit.

\renewcommand{\arraystretch}{1.3}
\begin{table}[htbp]\centering
\caption{The partial and total decay widths (in units of MeV) of
$2P$ states in the $B_s$ meson family. The forbidden decay channels
are marked by  --. For  $2P(1^+)/2P^\prime(1^+)$   states, we use
$\square$ to mark the allowed decay channels. \label{tab:Bs}}
\begin{tabular}{lcccccccccccc}
\toprule[1pt] Channels        &$2^3P_0$           &$2^3P_2$
& $2P(1^+)$/$2P^\prime(1^+)$       \\  \midrule[1pt]
$KB$            &65                 &2.6               &--                           \\
$KB^*$          & --                &7.3               &$\square$\\
$K^*B$          &--                 &1.6               &$\square$\\
$K^*B^*$        &36                 &69                &$\square$\\
$KB(1^3P_0)$    &--                 & --               &$\square$\\
$KB(1^3P_0)$    &--                 & --               &$\square$\\
$KB(1P(1^+))$   &5.3                &0.70              &$\square$\\
$KB(1P^\prime(1^+)$ &32                 &$3.5\times 10^{-3}$& $\square$\\
$\eta B_s$      &1.7                &$5.3\times 10^{-2}$&--\\
$\eta B_s^*$    &--                 &0.14              &$\square$\\
\midrule[1pt]
Total           &142                &82                &--\\
\bottomrule[1pt]
\end{tabular}
\end{table}

\begin{figure}[htbp]
\centering%
\begin{tabular}{c}
\scalebox{0.59}{\includegraphics{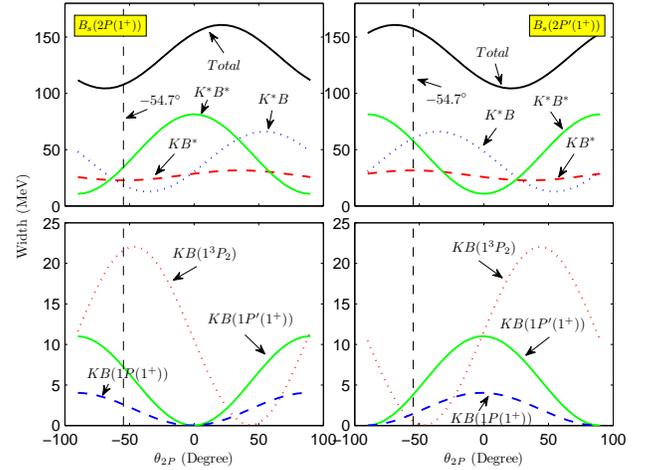}}
\end{tabular}
\caption{(color online). The variation of the decay width of
$B_s(1P(1^+))$ (the first column) and $B_s(1P^\prime(1^+))$ (the
second column) with the mixing angle $\theta_{2P}$. Here, we only
listed their main decay channels.}
 \label{bs2p}
\end{figure}

The decay behavior of the four $2P$ states are listed in Table
\ref{tab:Bs}, where $B_s(2P(1^+))$ and $B_s(2P^\prime(1^+))$ are the
mixed states satisfying Eq. (\ref{2pmix}). The three main decay channels for $B_s(2^3P_0)$ are $KB$, $K^*B^*$, and
$KB(1P^\prime(1^+))$. The sum of all partial decay widths listed in
the second column of Fig. \ref{bs2p} leads to the total decay width
around 142 MeV. Thus, $B_s(2^2P_0)$ is a broad $B_s$ meson. For
$B_s(2^3P_2)$, there is only one dominant decay channel $K^*B^*$,
where the branching ratio of $B_s(2^3P_2)\to K^*B^*$ is 0.84. At
present, $B_s(2^3P_0)$ and $B_s(2^3P_2)$ are still missing. Thus,
the obtained main decay modes of $B_s(2^3P_0)$ and $B_s(2^3P_2)$ may
be useful to the experimental search of $B(2^3P_0)$ and $B(2^3P_2)$.

The dependence of the total and partial decay widths of
$B_s(2P(1^+))$ and $B_s(2P^\prime(1^+))$ on $\theta_{2P}$ are
presented in Fig. \ref{bs2p}. If $\theta_{2P}$ takes the typical
value, the main decay modes are $B_s(2P(1^+))/B_s(2P^\prime(1^+))\to
K^{(*)}B^{(*)}$ (see Fig. \ref{bs2p} for detailed information).
We conclude that both $B_s(2P(1^+))$ and $B_s(2P^\prime(1^+))$ are
broad $B_s$ states,

\subsection{$2S$ and $3S$ states}

In the following, we illustrate the decay behavior of $B_s(2^1S_0)$,
$B_s(2^3S_1)$, $B_s(3^1S_0)$, and $B_s(3^3S_1)$ (see Table
\ref{bs1s2s}).

 \renewcommand{\arraystretch}{1.3}
\begin{table}[htbp]\centering
\caption{The partial and total decay widths (in units of MeV) of
$2S$ and  $3S$  states in the $B_s$ meson family. Here, we adopt --
to denote the forbidden decay channels. }\label{bs1s2s}
\begin{tabular}{lcccccccccccc}
\toprule[1pt] Channels          &$2^1S_0$ &$2^3S_1$&$3^1S_0$
&$3^3S_1$              \\  \midrule[1pt]
$KB$              &--       &17      &--                 &5.8                   \\
$KB^*$            &44       &34      &15                 &12                         \\
$K^*B$            &--       &--        &3.0                &0.41                  \\
$K^*B^*$          & --      &--       &21                 &12                      \\
$KB(1^3P_0)$         &--       &--      &$9.8\times 10^{-4}$&--                      \\
$KB(1^3P_2)$         &--       &--      &14                 &7.5                    \\
$KB(1P(1^+))$         &--       &--      &--                 &0.95                  \\
$KB(1P^\prime(1^+))$        &--       &--      & --                &6.0                   \\
$\eta B_s$        &--       &0.29    &--                 &0.17                  \\
$\eta B_s^*$      &0.14     &0.30    &0.28               &0.26                  \\
$\eta B_s(1^3P_0)$   &--       &--      &0.11               &--                     \\
$\eta B_s(1^3P_2)$   &--       &--      &$5.4\times 10^{-3}$&$1.2\times 10^{-2}$   \\
$\eta B_s(1P(1^+))$   &--       &--      & --                &0.15                   \\
$\eta B_s(1P^\prime(1^+))$  &--       &--      &--                 &$1.3\times 10^{-2}$    \\
$\phi B_s$        &--       &--      &0.26               &0.31
\\   \midrule[1pt]
Total             &44       &51      &54                 &46                        \\
\bottomrule[1pt]
\end{tabular}
\end{table}

According to the numerical results in  Table \ref{bs1s2s}, we
conclude:

\begin{itemize}
\item $B_s(2^1S_0)$ dominantly decays into $KB^*$. The contribution from
$\eta B_s(0^+)$ is negligible.

\item $KB^*$ and $KB$ are the two main decays of $B_s(2^3S_1)$.
The contribution from both $\eta B_s^*$ and $\eta B_s(0^+)$ decay
modes is quite small.

\item As the radial excitation of $B_s(2^1S_0)$, the total decay width of
$B_s(3^1S_0)$ is $\Gamma( B_s(3^1S_0))=54$ MeV. Here, the branching
ratios of $B_s(3^1S_0))\to KB^*, K^*B^*, KB(1^3P_2)$ are 0.28, 0.39,
and 0.26, respectively.

\item The main decay modes of $B_s(3^3S_1))$ include $KB^*$,
$K^*B^*$, $KB$, $KB(1^3P_2)$ and $KB(1P^\prime(1^+))$.

\end{itemize}

\subsection{$1D$ and $2D$ states}

In Table \ref{bs1d2d}, we list the numerical results of
$B_s(1^3D_1)$, $B_s(1^3D_3)$, $B_s(2^3D_1)$ and  $B_s(2^3D_3)$ and
the allowed decay channels of $B_s(1D(2^-))/B_s(1D^\prime(2^-))$ and
$B_s(2D(2^-))/B_s(2D^\prime(2^-))$.

\begin{table}[htbp]\centering
\caption{The partial and total decay widths (in units of MeV) of
$1D$ and $2D$ states in the $B_s$ meson family. The forbidden decay
channels are marked by  --. For the $1D(2^-)/1D^\prime(2^-)$ and
$2D(2^-)/2D^\prime(2^-)$ states, we use $\square$ to mark the
allowed decay channels. Here, the value $a \times 10^{-b}$ is
abbreviated as $a[b]$.}\label{bs1d2d}
\begin{tabular}{lcccccccccccc}
\toprule[1pt]
\multirow{2}{*}{Channels}     &\multirow{2}{*}{$1^3D_1$}     &  \multirow{2}{*}{$1^3D_3$}   &\multirow{2}{*}{ $2^3D_1$}           &\multirow{2}{*}{$2^3D_3$}           &$1D(2^-)$  &$2D(2^-)$    \\
&&&&  &  $1D'(2^-)$  &$2D'(2^-)$     \\   \midrule[1pt]
$KB$              &125    &5.2   &49            &0.61     &--           &--           \\
$KB^*$            &58     &5.7   &21            & 1.4[2]  &$\square$    &$\square$    \\
$K^*B$            &1.1    &3.0[5]&0.99          &5.5      &$\square$    &$\square$    \\
$K^*B^*$          &--     &--    &41            &13       &--           &$\square$    \\
$KB(1^3P_0)$         &--     &--    &--            & --      &--           &$\square$    \\
$KB(1^3P_2)$         &--     &--    &9.7           &4.3      &--           &$\square$    \\
$KB(1P(1^+))$         &--     &--    &12            &11       &--           &$\square$    \\
$KB(1P^\prime(1^+))$        &--     &--    &33            &0.38     &--           &$\square$    \\
$\eta B_s$        &4.8    &5.3[2]&1.7           & 1.2[2] &--           &--            \\
$\eta B_s^*$      &2.0    &4.5[2]&0.60          & 2.2[5]  &$\square$    &$\square$    \\
$\eta B_s(1^3P_0)$   &--     &--    &--            &  --     &--           &$\square$    \\
$\eta B_s(1^3P_2)$   &--     &--    &0.16          &5.8[2]  &--           &$\square$    \\
$\eta B_s(1P(1^+))$   &--     &--    &0.16          &0.14     &--           &$\square$    \\
$\eta B_s(1P^\prime(1^+))$  &--     &--    &0.13          &6.3[3] &--           &$\square$    \\
$\phi B_s$        &--     &--    &0.38          & 6.7[2] &--           &$\square$    \\
$\phi B_s^*$      &--     &--    &0.70          &1.7      &--
&$\square$    \\  \midrule[1pt]
Total             &191    &11    &171           &37       &--           &--                    \\
\bottomrule[1pt]
\end{tabular}
\end{table}

Our calculation shows

\begin{itemize}

\item Both $B_s(1^3D_1)$ and its radial excitation $B_s(2^3D_1)$ are
very broad, while both $B_s(1^3D_3)$ and $B_s(2^3D_3)$ are quite narrow.

\item The dominant decay modes of $B_s(1^3D_1)$ are $KB$ and $KB^*$ with the
branching ratios 0.65 and 0.30, respectively.

\item $B_s(2^3D_1)$ mainly decays into $KB$, $K^*B^*$,
$KB(1P^\prime (1^+))$, $KB^*$,  $KB(1P(1^+))$, and $KB(1^3P_2)$.

\item For $B_s(1^3D_3)$, there are two dominant decay modes,
$KB$ and $KB^*$. For $B_s(2^3D_3)$, $K^*B^*$, $KB(1P(1^+))$, $K^*B$, and
$KB(1^3P_2)$ are its main decay channels, while the remaining decay
channels have small partial decay widths (see Table \ref{bs1d2d} for
more details).

\end{itemize}

\begin{figure}[htbp]
\centering%
\begin{tabular}{c}
\scalebox{0.57}{\includegraphics{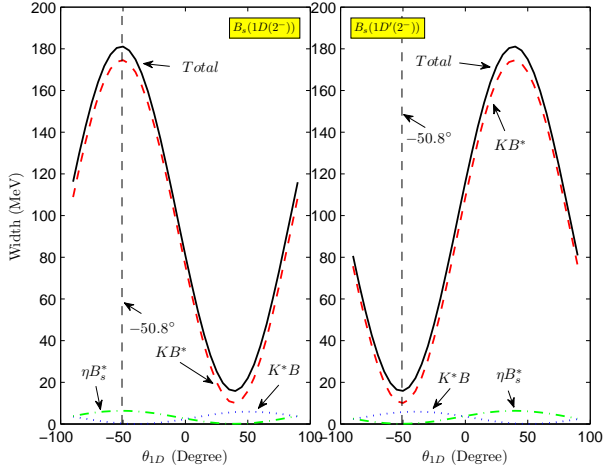}}
\end{tabular}
\caption{(color online). The dependence of the decay behavior of
$B_s(1D(2^-))$ (the first column) and $B_s(1D^\prime(2^-))$ (the
second column) on the mixing angle $\theta_{1D}$.}
 \label{bs1dmixing}
\end{figure}

\begin{figure}[htbp]
\centering%
\begin{tabular}{c}
\scalebox{0.57}{\includegraphics{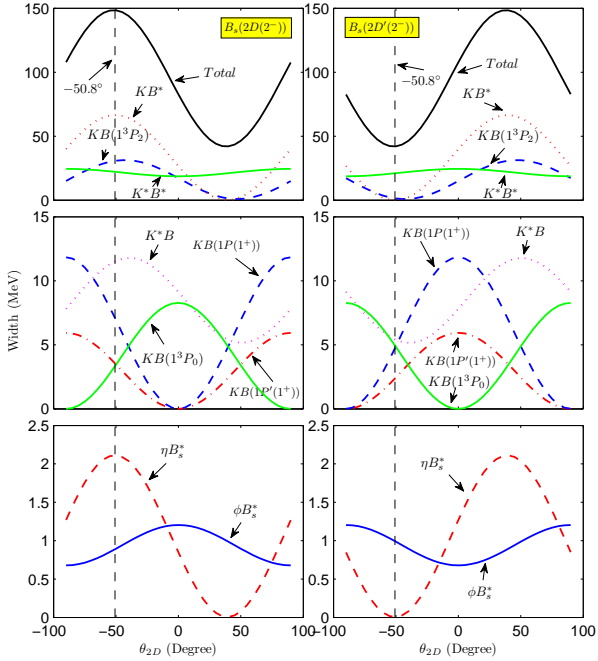}}
\end{tabular}
\caption{(color online). The dependence of the partial and total
decay widths of $B_s(2D(2^-))$ (the first column) and
$B_s(2D^\prime(2^-))$ (the second column) on the mixing angle
$\theta_{2D}$.}
 \label{bs2dmixing}
\end{figure}

As the mixed states $B_s(1D(2^-))/B_s(1D^\prime(2^-))$ and
$B_s(2D(2^-))/B_s(2D^\prime(2^-))$ are similar to
$B(1D(2^-))/B(1D^\prime(2^-))$ and $B(2D(2^-))/B(2D^\prime(2^-))$ in
Eq. (\ref{dmix}), in Figs. \ref{bs1dmixing} and
\ref{bs2dmixing} we thus show the variation of their partial and total
decay widths with $\theta_{1D}/\theta_{2D}$. With the typical
$\theta_{1D}=\theta_{2D}=-50.8^\circ$, we have the following
observations:
\begin{itemize}

\item For $B_s(1D(2^-))/B_s(1D^\prime(2^-))$, its $KB^*$ mode is
very important. The variation of the partial width
$B_s(1D(2^-))/B_s(1D^\prime(2^-))\to KB^*$  with the mixing angle is
very similar to that of the total decay width (see Fig.
\ref{bs1dmixing}).

\item $B_s(2D(2^-))$ is broad and mainly decays
into $KB^*$, $KB(1^3P_2)$, $K^*B^*$, and $K^*B$. As the parter of
$B_s(2D(2^-))$, the total decay width of $B_s(2D^\prime(2^-))$ is
smaller than that of $B_s(2D(2^-))$. The main decay mode of
$B_s(2D^\prime(2^-))$ is $K^*B^*$. While the sum of the partial
decay widths of the decay modes $KB(1P(1^+)), KB(1^3P_0),
KB(1P^\prime(1^+)), K^*B$ is comparable with the partial decay width
of $B_s(2D^\prime(2^-))\to K^*B^*$.

\end{itemize}

\subsection{$1F$ states}

\begin{figure}[htbp]
\centering%
\begin{tabular}{c}
\scalebox{0.57}{\includegraphics{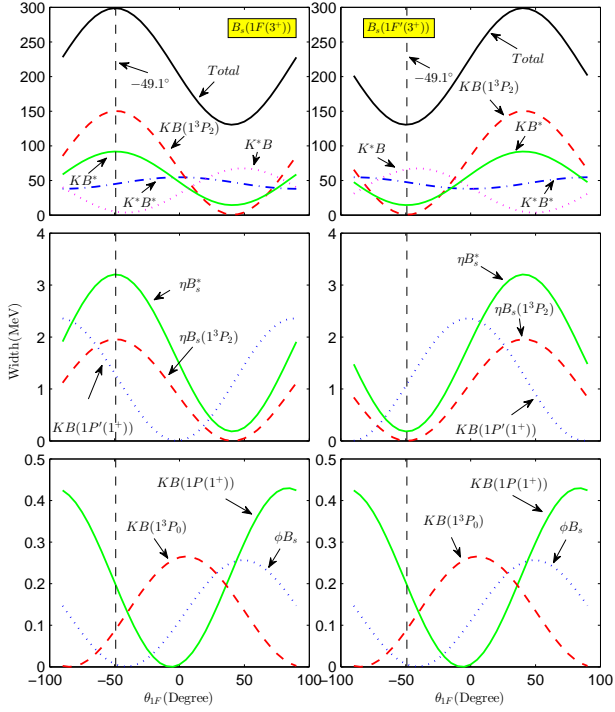}}
\end{tabular}
\caption{(color online). The variation of the partial and total
decay widths of $B_s(1F(3^+))$ (the first column) and
$B_s(1F^\prime(3^+))$ (the second column) with the mixing angle
$\theta_{1F}$.}
 \label{bs1fmixing}
\end{figure}

The decay behavior of the four $1F$ states is presented in Table
\ref{bs1f} and Fig. \ref{bs1fmixing}.

\begin{itemize}

\item $B_s(1^3F_2)$ is a very broad resonance. Its total decay width
can reach up to 319 MeV. The branching ratio of the
$KB(1P^\prime(1^+))$ mode is around 50\%. The other important decay
modes are $K^{(*)}B^{(*)}$ and $KB(1^3P_2)$.

\item $B_s(1^3F_4)\to K^*B^*$ dominates the total
decay width of $B_s(1^3F_4)$.

\item The total and partial decay widths of $B_s(1F(3^+))$ and
$B_s(1F^\prime(3^+))$ are dependent on the mixing angle
$\theta_{1F}$ in Eq. (\ref{fmix}). In Fig. \ref{bs1fmixing}, the
main and subordinate decay modes are given for $B_s(1F(3^+))$ and
$B_s(1F^\prime(3^+))$.

\end{itemize}

\begin{table}[htbp]\centering
\caption{The partial and total decay widths (in units of MeV) of
$1F$ states in the $B_s$ meson family. The forbidden decay channels
are marked by  --. For $1F(3^+)/1F^\prime(3^+)$  states, we use
$\square$ to mark the allowed decay channels. }\label{bs1f}
\begin{tabular}{lcccccccccccc}
\toprule[1pt] Channels        &$1^3F_2$           &$1^3F_4$
&$1F(3^+)$          \\  \midrule[1pt]

$KB$            &55                 &5.7                &--\\
$KB^*$          &38                 &7.4                &$\square$ \\
$K^*B$          &28                 &1.7                &$\square$\\
$K^*B^*$        &26                 &100                &$\square$\\
$KB(1^3P_0)$       &--                 & --                &$\square$\\
$KB(1^3P_2)$       &17                 &0.94               &$\square$\\
$KB(1P(1^+))$       &0.34               &0.55               &$\square$\\
$KB(1P^\prime(1^+))$      &149                &0.27               &$\square$\\
$\eta B_s$      &1.9                &$7.3\times 10^{-2}$&--\\
$\eta B_s^*$    &1.4                &$8.9\times 10^{-2}$&$\square$\\
$\eta B_s(1^3P_0)$ &--                 &--                 &$\square$\\
$\eta B_s(1^3P_2)$ &0.24               &$3.0\times 10^{-4}$&$\square$\\
$\eta B_s(1P(1^+))$ &$5.4\times 10^{-3}$&$3.8\times 10^{-4}$&$\square$\\
$\eta B_s(1P^\prime(1^+))$ &2.4               &$9.8\times 10^{-5}$&$\square$\\
$\phi B_s$      &0.14               &$7.1\times 10^{-5}$&$\square$\\
$\phi B_s^*$    &$2.3\times 10^{-3}$&--                 &$\square$\\
\midrule[1pt]
Total           &319                &116                &--\\
\bottomrule[1pt]
\end{tabular}
\end{table}

\begin{figure*}[htbp]
\centering%
\begin{tabular}{cc}
\scalebox{0.39}{\includegraphics{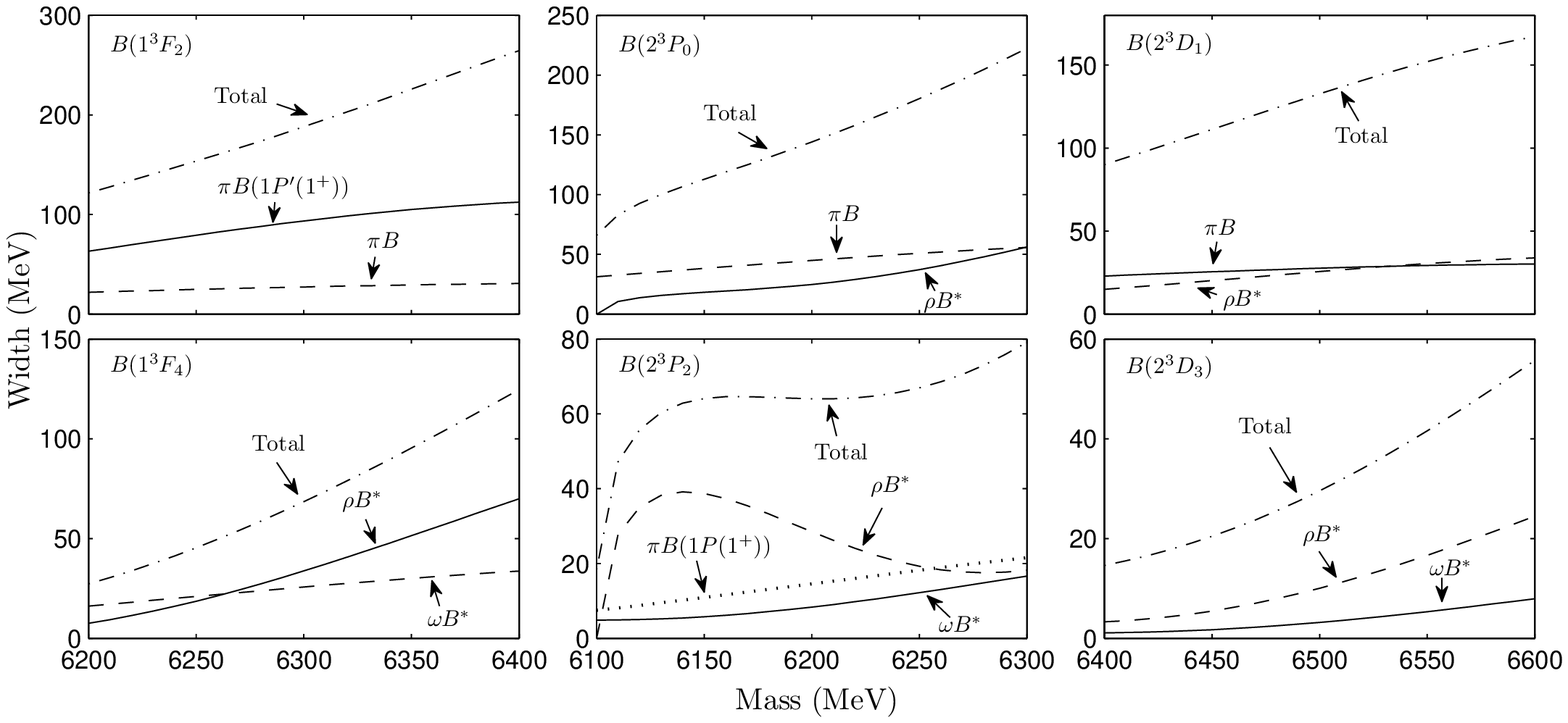}}&\scalebox{0.39}{\includegraphics{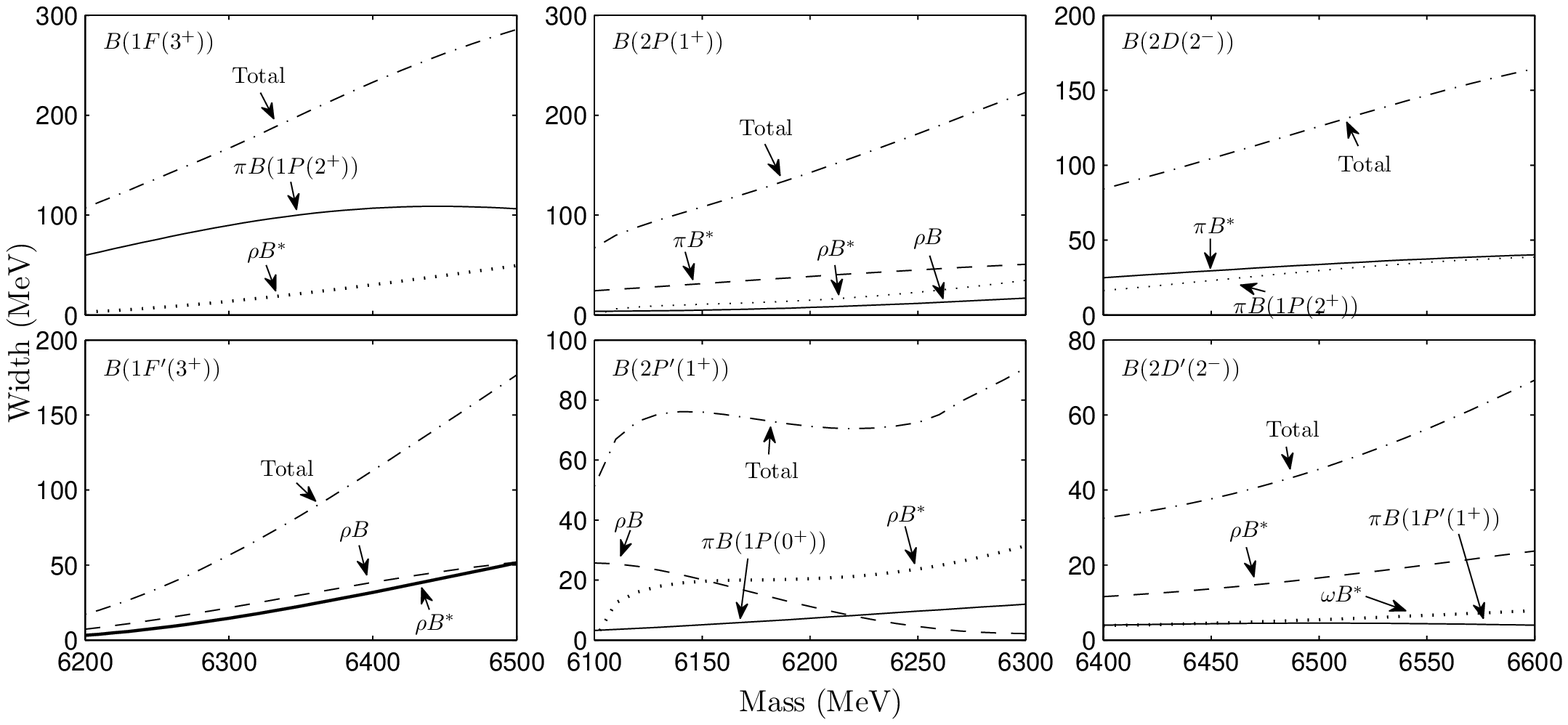}}\\
\scalebox{0.39}{\includegraphics{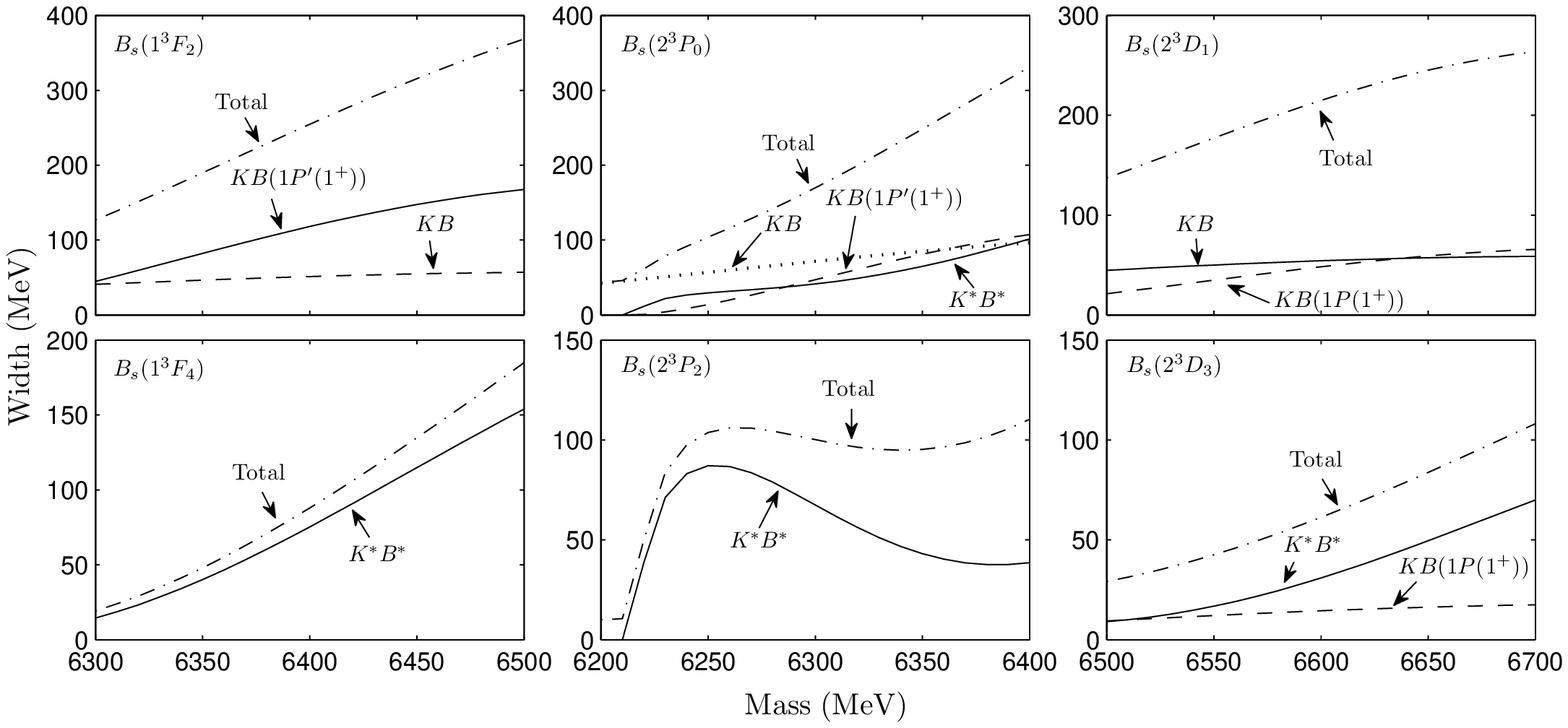}}&\scalebox{0.39}{\includegraphics{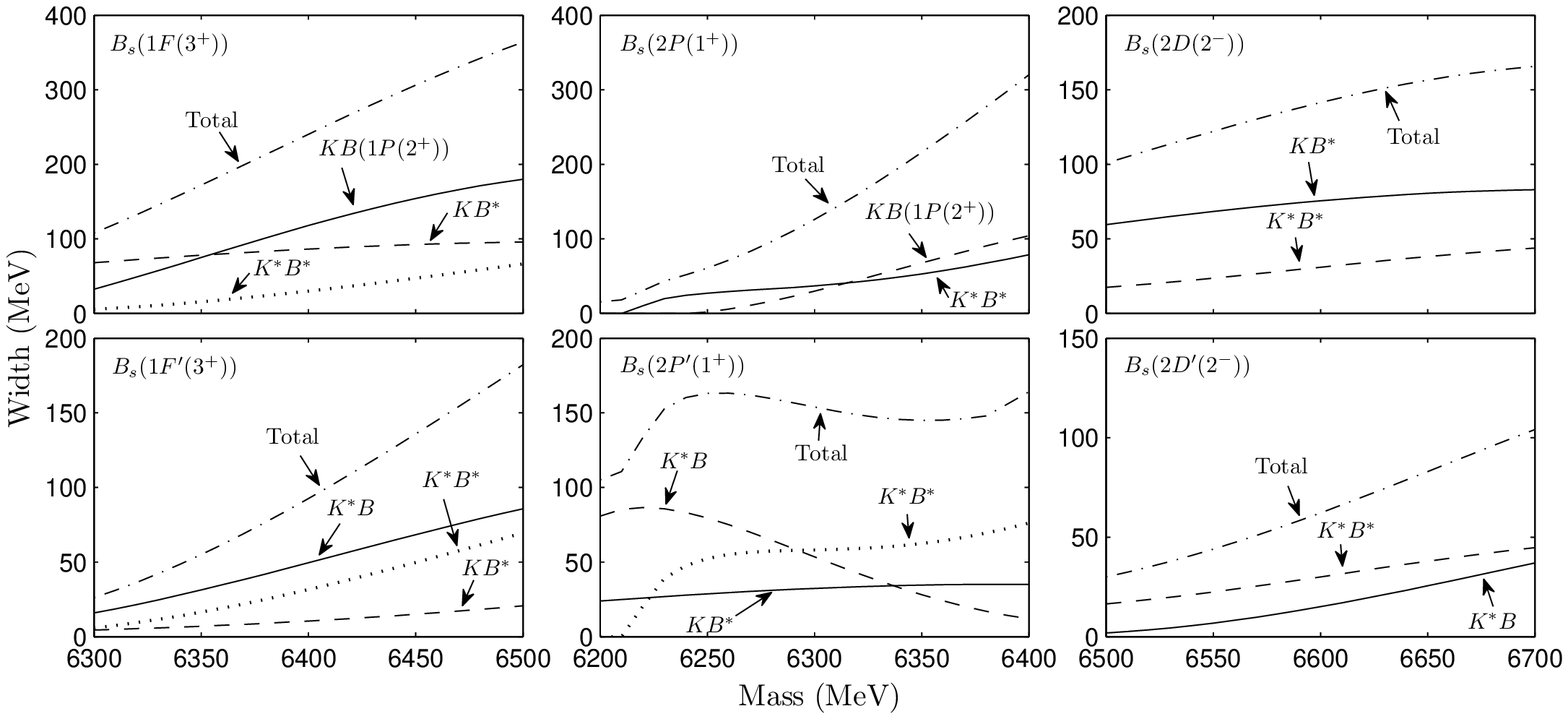}}
\end{tabular}
\caption{The variation of the decay widths (in units of MeV) of the higher $B$ and
$B_s$ mesons with their masses.}
 \label{massdependence}
\end{figure*}

In the above discussions, we take the predicted mass values of these
higher $B$ and $B_s$ meson families as input. However, the
uncertainty of the calculated masses sometimes is around $(50\sim
200)$ MeV from various theoretical models. Therefore, when
predicting the decay behavior of these higher $B$ and $B_s$ mesons,
we also study the variations of their decay behavior with the mass
of the parent state, which is illustrated in Fig.
\ref{massdependence}.

\section{summary}

Inspired by the recent experimental observation of the orbital
excitation $B(5970)$ for the first time by CDF Collaboration
\cite{Aaltonen:2013atp}, we have carried out a systematic study of
the higher $B$ and $B_s$ mesons. We have calculated both the masses
of the higher $B$ and $B_s$ mesons and their OZI-allowed two-body
strong decay patterns.

At present, the status of studying $B$ and $B_s$ mesons is very
similar to that of the $D$ and $D_s$ mesons in 2003. In the past
several years, CDF, D0, and LHCb Collaborations have played a very
important role in the study of the radial and orbital excitations of
the $B$ and $B_s$ meson families. The higher radial and orbital
excitations of the $B$ and $B_s$ mesons begin to emerge in
experiment. In the coming years, LHCb has the potential to discover
more and more excited $B$ and $B_s$ mesons. Hopefully
our present investigation will be helpful to future experimental
searchs for these interesting heavy mesons.

%\vfil
\section*{Acknowledgements}

This project is supported by the National Natural Science Foundation
of China under Grants No. 11222547, No. 11175073, No. 11035006, No. 11375240 and No. 11261130311,
the Ministry of Education of China (FANEDD under Grant No. 200924,
SRFDP under Grant No. 2012021111000, and NCET), and the Fok Ying Tung
Education Foundation (Grant No. 131006).

\vfil

\end{document}